\documentclass[%
 preprint,
 superscriptaddress,
 groupedaddress,
 showpacs,
 amsmath,amssymb,
 aps,
]{revtex4-1}
\usepackage{color,soul}
\usepackage{graphicx}
\usepackage{dcolumn}
\usepackage{bm}

\begin{document}


\title{Revisiting Submicron-Gap Thermionic Power Generation Based on Comprehensive Charge and Thermal Transport Modeling}

\author{Devon Jensen}
\author{Mohammad Ghashami}
\affiliation{Department of Mechanical Engineering, University of Utah, Salt Lake City, Utah 84112, United States.}
\author{Keunhan Park}
\email{kpark@mech.utah.edu}
\affiliation{Department of Mechanical Engineering, University of Utah, Salt Lake City, Utah 84112, United States.}

\begin{abstract}
Over the past years, thermionic energy conversion (TEC) with a reduced inter-electrode vacuum gap has been studied as an effective way to mitigate a large potential barrier due to space charge accumulation. However, existing theoretical models do not fully consider the fundamental aspects of thermionic emission when the inter-electrode gap shrinks to the nanoscale, which results in underestimation of thermionic power generation for such small gaps. The present work addresses this challenge by comprehensively modeling charge and thermal transport processes with specific consideration of nanoscale gap effects, such as image charge perturbation, electron tunneling, and near-field thermal radiation. Carefully conducted energy balance analysis reveals that if optimized, submicron-gap TEC can excel the micron-gap counterpart with $\sim$4 times the power output and $\sim 5-10$ \% higher energy conversion efficiency. Moreover, the high-temperature collector of the submicron-gap TEC, which is due to thermionic and near-field radiative heat transfer, allows the addition of a bottom-cycle heat engine to further enhance the power and efficiency when combined. Electric field concentration due to engineered surface roughness is also examined as a potential approach to produce an additional increase in power generation. We believe that the present work provides a theoretical framework for submicron-gap thermionic power generation as a promising energy recycling scheme for high-quality heat sources.

\end{abstract}

\maketitle


\section{\label{sec:level1}Introduction}
Thermionic energy conversion (TEC) is a class of direct heat-to-electrical power generation systems, configured with a thermionic emitter (or cathode) and collector (or anode) separated by a vacuum space \cite{Langmuir1923,Hatsopoulos1973}. When the emitter is heated to a high temperature, typically over 1400 K \cite{Hatsopoulos1973}, electrons with higher energies than the emitter's work function are evaporated from the surface. When the electrodes of a thermionic cell are connected to an electrical load, the emitted electrons can be swept to the collector to generate electric power \cite{Lee2014}. The theoretical efficiency limit of TEC is higher than those of thermophotovoltaic and thermoelectric heat engines at high heat source temperatures \cite{Shakouri2005}. In addition, photovoltaic and thermoelectric schemes have technical challenges for reliable high-temperature operations, such as mitigating a large dark current in photovoltaic cells and maintaining a large temperature gradient in thermoelectic cells \cite{Hishinuma2001,Skoplaki2009,Lee2014,Zhu2015}. 

Despite several appealing features, TEC has not been widely adopted due to poor energy conversion processes in reality even at high temperatures: the highest efficiency obtained with lab-demonstrated thermionic energy conversion is still below 10\% \cite{Yuan2017}. Such low efficiency first stems from high emitter and collector work functions. For example, the work function of tungsten, routinely used for thermionic electrodes, is $\sim$4.5 eV \cite{Koeck2012} and should be reduced to obtain sufficient thermionic power output and conversion efficiency. Over the years, several methods have been proposed to lower work functions, such as barium surface coatings \cite{Littau2013}, atomic surface modification \cite{Suzuki2009,Voss2014,Kato2016}, and surface nano-texturing \cite{Barmina2012}. However, when the electrostatic field between the electrodes is not strong enough to immediately attract emitted charges to the collector, lower work function electrodes do not necessarily improve thermionic performance. Under this condition, electrons accumulate between the electrodes, inducing a repulsive force that increases a potential barrier for electron emission  \cite{Hatsopoulos1973, Hatsopoulos1979, Lee2012j, Jensen2017b, Lim2018}. This negative space charge effect has been mitigated by filling the vacuum space with positive ions \cite{Hatsopoulos1973, Xiao2017}. However, a continuous supply of positive ions and the scattering of electrons upon collisions with ions and neutralized atoms can lower the overall system efficiency. Although Meir \textit{et al.} \cite{Meir2013a} proposed inserting a positively biased micro-mesh gate between the electrodes to accelerate electrons to the collector, fabricating the gate structure between the emitter and collector is not an easy task, and a magnetic field should be applied parallel to the averaged electron trajectories to reduce the current lost to the gate. A recent study proposed using two-dimensional (2D) materials such as graphene for magnetic-field-free gate electrodes \cite{Wanke2016}, but this concept is not ready to be realized due to technical challenges in large-scale synthesis and nanofabrication of 2D materials.  

Another feasible approach to mitigate the space-charge effect is to reduce the vacuum gap distance between the emitter and collector. Hatsopoulos and Gyftopoulos \cite{Hatsopoulos1973} claimed that reducing the inter-electrode vacuum gap can suppress space-charge accumulation by increasing the electric field in the vacuum space. Since then, previous works have demonstrated potential benefits of reducing the inter-electrode vacuum gap on thermionic power generation \cite{Hishinuma2001, Zeng2006, ODwyer2009a, Lee2009f, Lee2012j,Voss2013,Wang2016}. However, each model lacks one or more critical aspects of reduced gap effects in thermal and charge transport processes, and  has not fully considered energy balance within a TEC system. Such lacking aspects in modeling has led to the underestimation of thermionic power output and conversion efficiency for submicron-ranged vacuum gaps, suggesting that the vacuum gap distance should be in the single-digit micrometer range for optimal thermionic power generation in the typical emitter temperature range at around 1500 K \cite{Lee2012j, Littau2013, Lee2014, Lim2018}.

In the present article, we report a theoretical study that challenges the long-believed design principle of a TEC system by demonstrating compelling advantages of further reducing the inter-electrode vacuum gap to the sub-micrometer regime. Charge and thermal transport processes across a reduced vacuum gap are rigorously modeled by considering gap effects, such as image-charge potentials, electron tunneling, and near-field thermal radiation, in a comprehensive manner. Systematic energy balance analysis is the key to understand the role of thermionic emission and thermal radiation in overall energy transfer across the vacuum gap, from which the thermionic power output and energy conversion efficiency can be accurately calculated. The obtained results reveal that if optimized, a sub-micrometer vacuum gap permits the best performance of a vacuum TEC system. Moreover, our work investigates possibilities to further improve the TEC performance by topping a TEC device in a combined cycle configuration and concentrating local electric fields on nano-engineered electrode surfaces. 

\section{\label{sec:level1}Modeling}
\subsection{\label{Thermionic} Charge transport processes}
\begin{figure}[t!]
    \centering\includegraphics[width=0.7\linewidth]{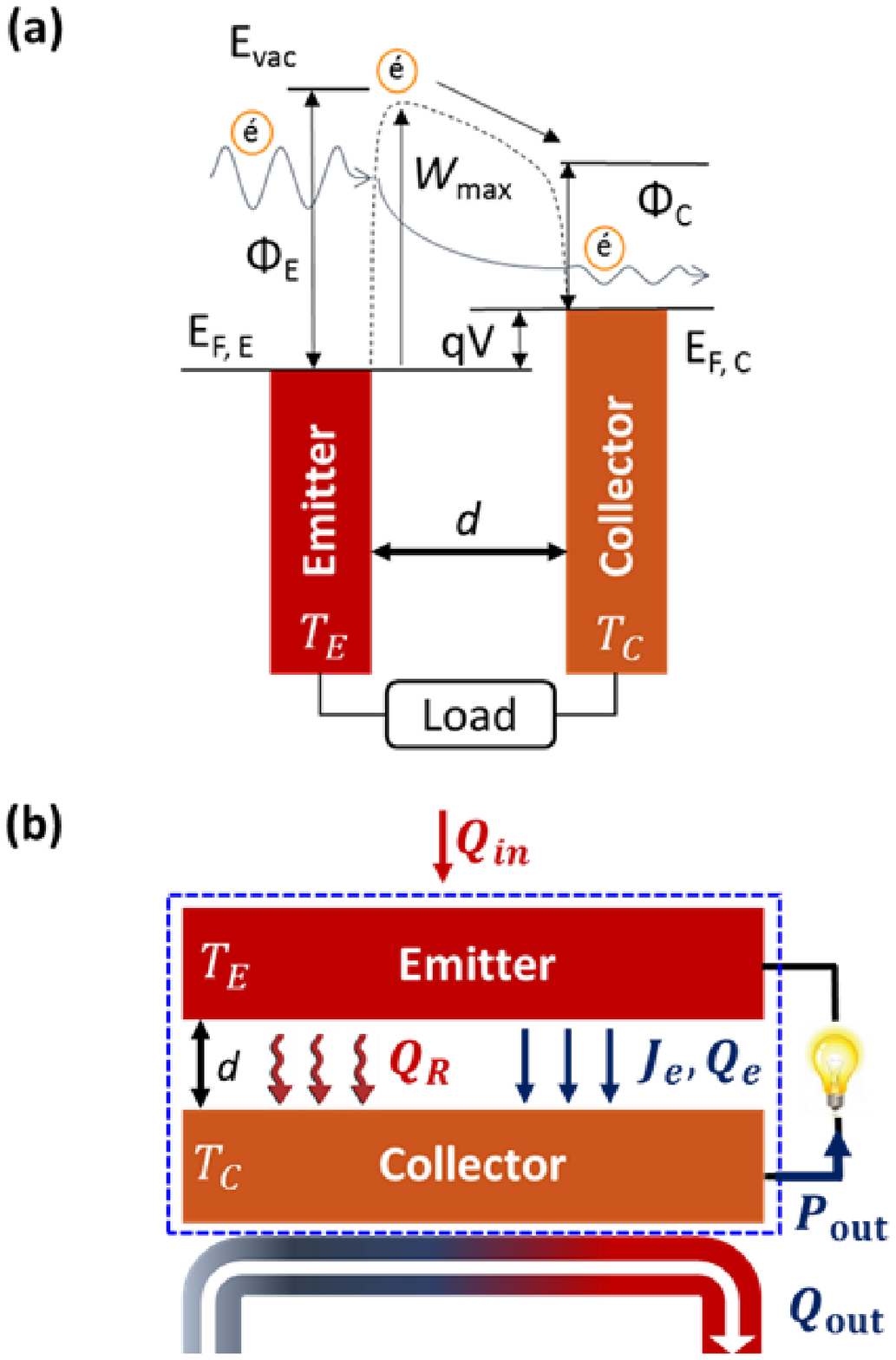}
    \caption{}
    \label{Fig1}
\end{figure}

Fig. \ref{Fig1}(a) illustrates the schematics and energy diagram of a submicron-gap vacuum TEC system that consists of a thermionic emitter with a work function $\Phi_E$ and a collector with a work function $\Phi_C$, connected to a electrical load that causes a voltage drop $V$. When the emitter temperature is elevated to $T_E$, electrons can be emitted and travel to the collector at temperature $T_C$ by thermionic emission and quantum electron tunneling phenomena. The net current density carried by electrons across the inter-electrode gap can be calculated as $J_e = J_{\mathrm{TE}} + J_{\mathrm{QE}}$, where $J_{\mathrm{TE}}$ is the net thermionic current density and $J_{\mathrm{QE}}$ is the net quantum electron tunneling current density. In general, the current density from electrode $i$ can be expressed as \cite{ODwyer2005a,Jensen2017b}
\begin{equation}
    \label{Eqn:Ji}
    J_i = q\int_{}^{}\mathcal{T}(E_x)N_i(E_x,T_i)dE_x
\end{equation}
where $i$ is either the emitter ($E$) or the collector ($C$), $q$ is the electron charge, $\mathcal{T}(E_x)$ is the electron transmission probability, and $E_x$ is the electron energy normal to the surface (i.e., $E_x=m_e v_x^2/2$ with $m_e$ and $v_x$ being the free electron mass and the electron velocity component normal to the surface, respectively). $N_i(E_x,T_i)$ is the supply function of electrode $i$, denoting the number of electrons at a normal energy level $E_x$ and temperature $T_i$ per unit area and unit time. When electrons have $E_x$ higher than the maximum potential barrier $W_\textrm{max}$, they are thermionically emitted from the electrode and travel through the vacuum space with a transmission probability of $\mathcal{T}(E_x) = 1$. In addition, high electron energies above the Fermi level (i.e., $E_{F,i}$ in Fig. \ref{Fig1}(a)) allow electrons to be approximated by a Maxwell-Boltzmann distribution, yielding $N_i(E_x,T_i) = (m_e k_B T_i/2 \pi^2 \hbar^3)\textrm{exp}[-(E_x-E_{F,i})/k_B T_i]$ with $k_B$ and $\hbar=h/2\pi$ being the Boltzmann constant and the reduced Planck constant, respectively \cite{Jensen2017b}. $J_{\mathrm{TE}}$ is then derived by integrating Eq. (\ref{Eqn:Ji}) from $E_x=W_\textrm{max}$ to infinity to yield the Richardson-Dushman equation:
\begin{equation}
    \label{Eqn:Jte}
    J_{\textrm{TE}} =  J_{\textrm{TE},E} -     J_{\textrm{TE},C} =  AT_{E}^2\textrm{exp}\left[\frac{-W_{\textrm{max}}}{k_BT_{E}}\right] - AT_{C}^2\textrm{exp}\left[\frac{-(W_{\textrm{max}}-qV)}{k_BT_{C}}\right]
\end{equation}
where $A = 4\pi m_e k_B^2q/h^3$ is the Richardson constant. When the inter-electrode vacuum gap is reduced to the nanoscale, electrons can tunnel through the vacuum gap although their energy is below $W_\textrm{max}$. For quantum electron tunneling, the transmission probability is no longer unity but approximated as $\mathcal{T}(E_x) = {\mathrm {exp}}[-\theta(E_x)]$, where $\theta(E_x)$ can be written as the following equation based on the  Wentzel-Kramers-Brillouin (WKB) approximation \cite{Murphy1956,Christov1966a,Hishinuma2001,Jensen2003, Lee2009f,Wang2016,Baeva2018}:
\begin{equation}
    \label{Eq_theta}
    \theta(E_x) = \frac{\sqrt{8m_e}}{\hbar}\int_{x_1}^{x_2} \sqrt[]{W(x)-E_x}dx
\end{equation}
Here, $x_1$ and $x_2$ are the locations at which $E_x$ equals the local potential barrier $W(x)$, designating the width of the electron tunneling barrier at $E_x$. In addition, the Fermi-Dirac distribution should be used for the supply function, yielding $N_i(E_x,T_i) = (m_e k_B T_i/2 \pi^2 \hbar^3) \textrm{ln}\{1+\textrm{exp}[-(E_x-E_{F,i})/k_B T_i]\}$ for electron tunneling \cite{Jensen2003}. Therefore, the net electron tunneling current density can be written as 
\begin{equation}
    \label{Eqn:Jqe}
    \begin{split}
        J_{\textrm{QE}} & = J_{\textrm{QE},E} -     J_{\textrm{QE},C} \\
        & =  q\int_{-\infty}^{W_{\textrm{max}}}\mathcal{T}(E_x)\left[N_E(E_x,T_E) - N_C(E_x-qV,T_C)\right]dE_x
    \end{split}
\end{equation}
which should be numerically calculated. 

For the accurate calculation of the net current density, it is crucial to obtain the potential barrier profile in the vacuum gap $W(x)$ and determine $W_{\textrm{max}}$ from the obtained profile. Within the electrostatic framework, the potential barrier profile can be written as
\begin{equation}
    \label{Eqn:W(x)}
    W(x) = W_{\textrm{id}}(x) + W_{\textrm{sc}}(x) + W_{\textrm{ic}}(x)
\end{equation}
Here, $W_{\textrm{id}}(x)$ is the ideal barrier profile while $W_{\textrm{sc}}(x)$ and $W_{\textrm{ic}}(x)$ denote space-charge and image-charge perturbations, respectively. Under the ideal condition, the potential barrier shows a linear profile in terms of the electrode work functions ($\Phi_E$ and $\Phi_C$) and the load voltage $V$  \cite{Simmons1963a,Baldea2012a}: 
\begin{equation}
    \label{Eqn6}
    W_{\textrm{id}}(x) = \Phi_E - (\Phi_E-\Phi_C-qV)\left(\frac{x}{d}\right)
\end{equation}
where $d$ is the gap distance, and $x$ is the position between the emitter surface ($x=0$) and the collector surface ($x=d$). 
As electrons accumulate in the vacuum space, negative charges are built up to impede further emission of electrons from the emitter surface. $W_{\textrm{sc}}(x)$ represents the effect of this negative space charge accumulation on the potential barrier profile. Under the assumption of collisonless electron transport, $W_{\textrm{sc}}(x)$ is calculated by numerically solving Poisson`s equation to satisfy the half-Maxwellian distribution of electron velocities at the $W_\textrm{max}$ position in the vacuum space (i.e., $dW(x_\textrm{max})/dx = 0$ for $0 < x_\textrm{max} < d$) \cite{Langmuir1923, Hatsopoulos1973, Hatsopoulos1979, Lee2012j, Jensen2017b}.
On the other hand, $W_{\textrm{ic}}(x)$ can be calculated by accounting for electrostatic interactions between image charges in both electrodes due to the presence of electrons in the vacuum gap \cite{Baldea2012a}: 
\begin{equation}
    \label{Eqn:W_ic}
    \begin{split}
        W_{\textrm{ic}}(x) = \frac{q^2}{16\pi \epsilon_0 d}\left[-2\Psi(1)+\Psi\left( \frac{x}{d}\right)+\Psi\left(1-\frac{x}{d}\right)\right]
    \end{split}
\end{equation}
where $\epsilon_0$ is the permittivity of the free space and $\Psi=d\mathrm{log}\Gamma(x)/dx$ is the digamma function. The above equation is essentially identical to the image potential equation in other works \cite{Simmons1963a,Hishinuma2001,Wang2016}. It should be noted that for macroscale vacuum gaps, $W_{\textrm{sc}}(x)$ is the dominant factor that increases $W(x)$ (and subsequently $W_{\textrm{max}}$) while $W_{\textrm{ic}}(x)$ is negligibly small. However, as the vacuum gap decreases, the electric field between the electrodes become stronger to suppress the space charge effect. Meanwhile, a small distance between real and image charges further augments the electric field, which lowers $W(x)$ below the ideal profile particularly near the electrode surfaces \cite{Jensen2017b}. 

\subsection{\label{Heat Transfer} Energy Balance Analysis}
As illustrated in Fig. \ref{Fig1}(b), energy transfer by electrons ($Q_e$) is partly used to generate power ($P_{\mathrm{out}}$) while the remainder is converted to heat in the collector. In addition, there is radiative heat transfer ($Q_R$) across the vacuum gap that heats the collector. The thermionic energy conversion efficiency can be expressed as 
\begin{equation}
    \eta = \frac{P_\mathrm{out}}{Q_{\textrm{in}}} = \frac{P_{\textrm{TE}}+P_{\textrm{QE}}}{Q_e+Q_{\textrm{R}}}
    \label{Eqn15}
\end{equation}
where $Q_{\textrm{in}}$ is the heat input to the emitter and should be balanced with  $Q_e+Q_{\textrm{R}}$ to maintain the emitter  temperature at the steady state. The thermionic power output can be calculated by $P_\mathrm{out} = P_\textrm{TE} + P_\textrm{QE}=[V( J_{\textrm{TE}}+J_{\textrm{QE}})]_{\textrm{max}}$, where $V_{\mathrm{max}}$ is the operational voltage at the maximum power output. The electron-carried energy flux across a submicron-sized vacuum gap includes thermionic emission and electron tunneling, i.e., $Q_e=Q_{\mathrm{TE}}+Q_{\mathrm{QE}}$, where 
\begin{equation}
        Q_{\textrm{TE}} =\frac{1}{q}\left[J_{\textrm{TE}}W_{\textrm{max}} + 2k_B\left(J_E T_E - J_C T_C\right)\right]
    \label{Eqn:Q_TE}
\end{equation}
is the thermionic energy flux \cite{Lee2012j}. Here, the second term denotes the kinetic energy carried away from each electrode by electrons \cite{Hatsopoulos1979}. The electron tunneling energy flux can be calculated by \cite{Hishinuma2001,Lee2009f} 
\begin{equation}
\begin{split}
        Q_{\textrm{QE}} = \int_{-\infty}^{W_{\textrm{max}}}\mathcal{T}(E_x)&[(E_x + k_B T_E) N_E(E_x,T_E) \\
        & - (E_x + k_B T_C)N_C(E_x-qV, T_C)]dE_x
\end{split}
\label{Eqn:Q_QE}
\end{equation}
The radiative heat transfer across the vacuum gap under study should have near-field enhancement due to photon tunneling of thermally excited evanescent electromagnetic waves \cite{Park2013}. The near-field radiative heat flux from the emitter to the collector can be calculated by the following equation \cite{Joulain2005c,Basu2009,Park2013}: 
\begin{equation}
    \label{Eqn13}
         Q_{\textrm{R}} = \frac{1}{\pi^2}\int_{0}^{\infty}d\omega[\Theta(\omega,T_E)-\Theta(\omega,T_C)] \int_{0}^{\infty}Z(k_\parallel,\omega)k_\parallel dk_\parallel
\end{equation}
where $\Theta(\omega,T_i)=\hbar \omega/[\mathrm{exp}(\hbar\omega/k_B T_i)-1]$ is the mean energy of a Planck oscillator at angular frequency $\omega$, and $k_\parallel$ is the wavevector component parallel to the surface. $Z(k_\parallel,\omega)$ is the exchange function that can be formulated by the dyadic Green's function within the fluctuational electrodynamics framework. The detailed formulation of $Z(k_\parallel,\omega)$ for a three-layer configuration with a semi-infinite emitter and collector can be found in previous works \cite{Joulain2005c,Basu2009,Park2013} and will not be repeated here. Heat rejection from the collector to maintain $T_C$ is simply modeled as $Q_\textrm{out} = h_\infty(T_\textrm{C} - T_\infty)$, where $h_\infty$ is the convection heat transfer coefficient. $T_\infty$ is the environmental temperature and set to 300 K to evaluate the overall heat transfer rate to the environment. 

In the present study, the TEC system is configured with tungsten-barium-oxygen electrodes \cite{Jacobs2017}. The emitter temperature is assumed to be at $T_E=1575$ K, which is a routine operational configuration of commercial dispenser cathodes for thermionic emission. The work function of each electrode is optimized to satisfy $\Phi = T/750$ to secure the best performance of the TEC system \cite{Hatsopoulos1973}, which yields $\Phi_E=2.10$ eV for the emitter. Two design-point scenarios for the collector side are considered: a constant collector temperature at $T_C=1000$ K ($\Phi_C=1.33$ eV) and a constant heat transfer coefficient at $h_\infty=1000$ W/m$^2$-K for heat rejection. The dielectric functions of the emitter and the collector are calculated from the Drude model, which can be written as $\epsilon (\omega, T_i) = 1 - \sigma_{0,i}/[\tau_i\epsilon_0(\omega^2+i\omega/\tau_i)]$. Here, $\sigma_{0,i}$ is the DC conductivity \cite{VanDerMaas1985} and $\tau_i$ is the electron relaxation time of electrode $i$ \cite{Roberts1959,Lee2012j}, both of which are considered to be temperature-dependent.

\section{\label{sec:level1}Results and Discussion}
\begin{figure}[t!]
\centering\includegraphics[width=1\linewidth]{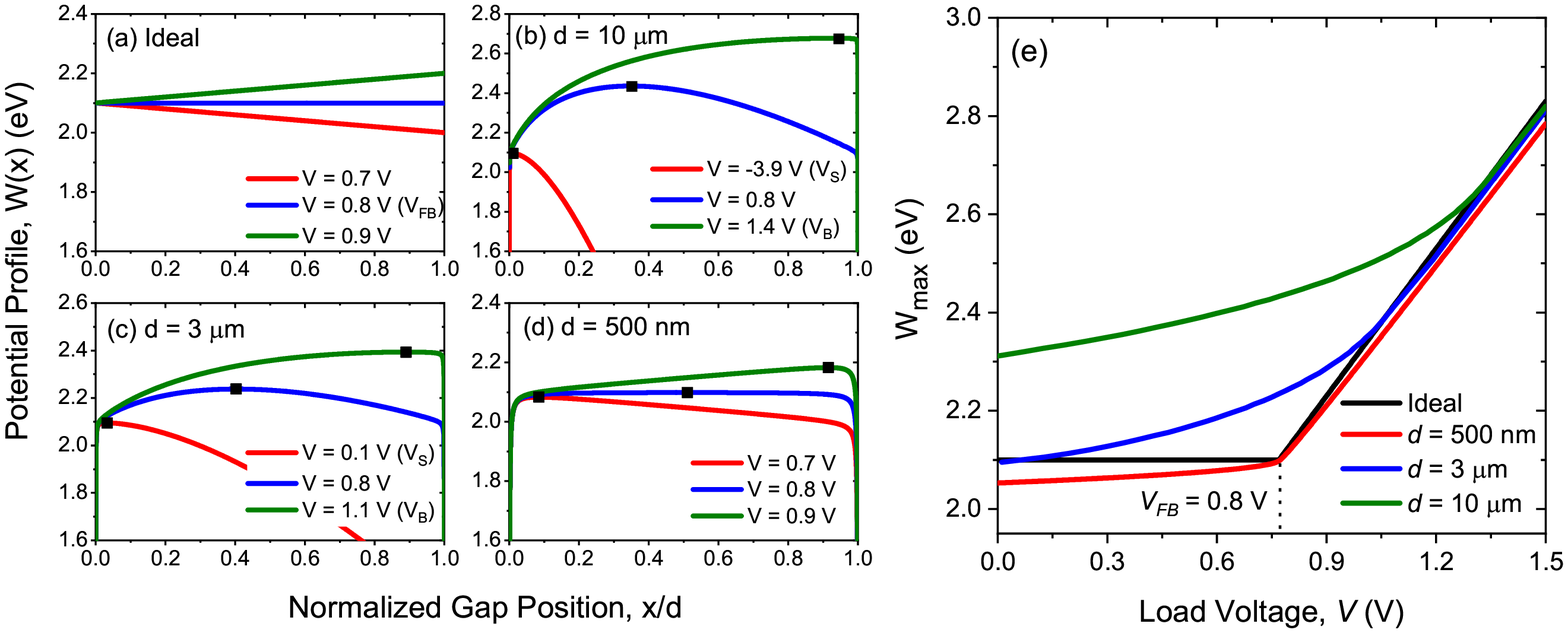}
\caption{}
\label{Fig2}
\end{figure}

As mentioned in the modeling section, accurate calculation of the potential profile $W(x)$ is crucial for the reliable performance analysis of thermionic power generation. Fig. \ref{Fig2} shows $W(x)$ as a function of the normalized gap position, $x/d$, and the corresponding $W_{\mathrm{max}}$ as a function of the load voltage for different gap distances. For the ideal case with $\Phi_E = 2.10$ eV and $\Phi_C = 1.33$ eV, $W(x)$ exhibits a linear profile across the vacuum gap: see Fig. \ref{Fig2}(a). As shown in \ref{Fig2}(e), the resulting $W_{\textrm{max}}$ remains constant at $\Phi_E$ as the load voltage increases to the flat-band voltage (i.e., $V_{\mathrm{FB}}=(\Phi_E-\Phi_C)/q=0.8$ V), and linearly increases with further increase of the load voltage (i.e., $W_{\textrm{max}}=\Phi_C+qV$). However, when the thermionic electrodes are separated by microscale gap distances as shown in Figs. \ref{Fig2}(b) and (c), the accumulation of negative charges in the vacuum space causes a parabolic-like potential energy profile with $W_{\textrm{max}}$ greater than the ideal case. Besides the flat-band voltage, potential profiles at the Boltzmann voltage ($V_B$) and the saturation voltage ($V_S$) are also shown in Figs. \ref{Fig2}(b) and (c) to better describe the dependence of the potential profile on the load voltage. For $V>V_B$, charge transport is severely impeded by a high potential barrier near the collector surface to generate insignificant thermionic power. When the load voltage is applied below the saturation voltage limit (i.e., $V<V_S$), electrons are accelerated by the electric field across the vacuum gap to yield $W_{\textrm{max}}$ smaller than the ideal case (i.e., $W_\mathrm{max} = \Phi_E$).  However, $V_S$ is typically in the negative voltage range for microscale gaps, making the acceleration regime inadequate for thermionic power generation. When the vacuum gap is in the sub-micrometer range as shown in Fig. \ref{Fig2}(d), the potential profiles become similar to the ideal profiles due to the suppression of the space charge effect. Moreover, the image charge effect accelerates electrons in the entire load voltage range, further reducing $W_\mathrm{max}$ below the ideal curve: see $d=500$ nm in Fig. \ref{Fig2}(e).

\begin{figure}[t!]
    \centering\includegraphics[width=0.8\linewidth]{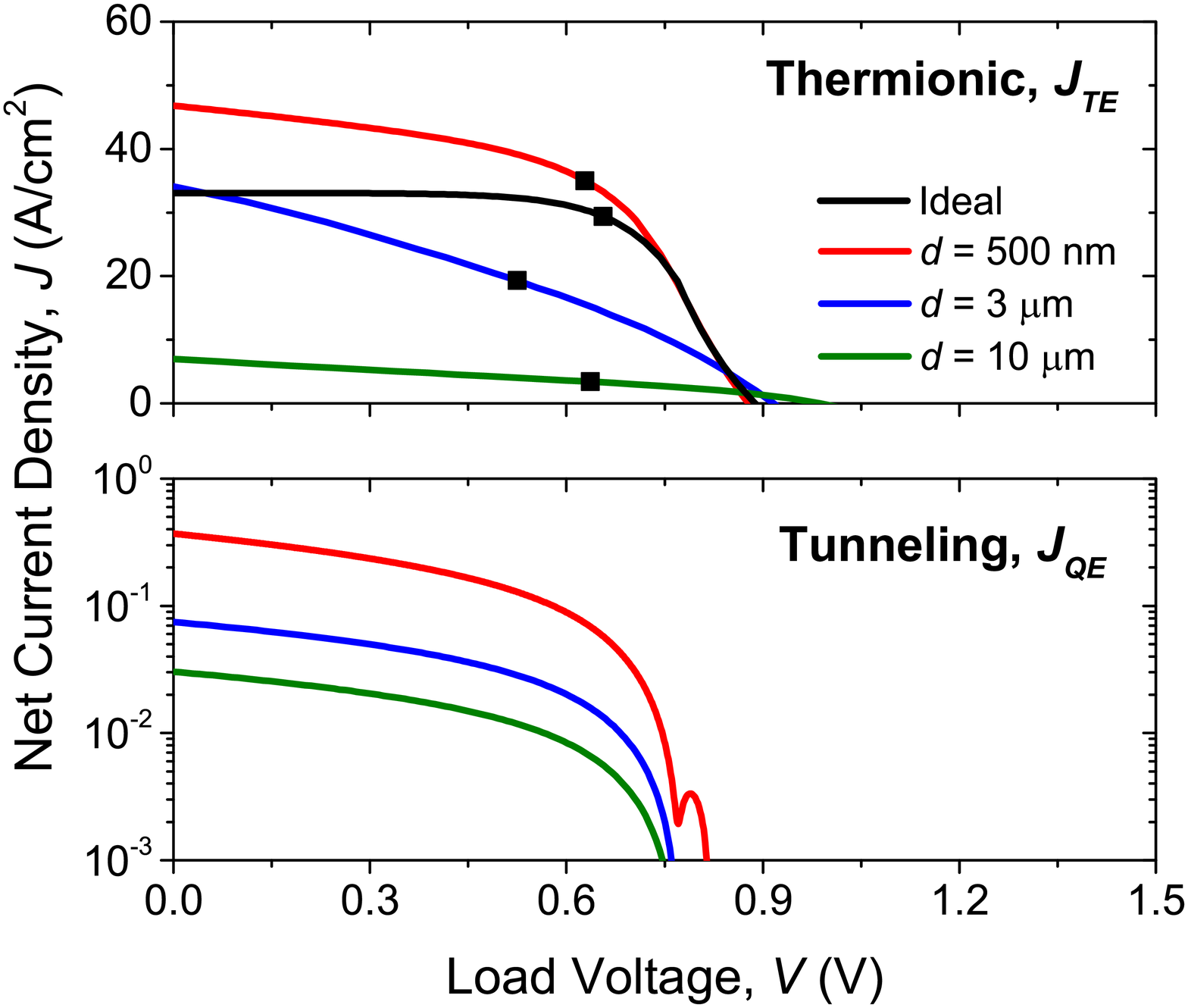}
    \caption{}
    \label{Fig3}
\end{figure}

Based on the potential profile and $W_\textrm{max}$, the net thermionic and tunneling current densities ($J_\mathrm{TE}$ and $J_\mathrm{QE}$, respectively) are calculated and plotted in Fig. \ref{Fig3}. The ideal $J_\mathrm{TE}$ curve is flat up to $V \approx 0.75 V_\mathrm{FB}$, and exponentially decays for higher load voltages. For microscale gap distances (e.g., $d=3$ $\mu$m and 10 $\mu$m), the current density values are significantly diminished due to the negative space charge effect. In contrast, a submicron vacuum gap (i.e., $d=500$ nm) generates a higher current density than the ideal $J$-$V$ curve due to the image charge effect, which is consistent with $W_\mathrm{max}$ in Fig. \ref{Fig2}(e). The net tunneling current density exponentially decays as the load voltage increases, with a steeper slope at higher voltages, while increasing as $d$ shrinks into the submicron range. However, its contribution to total power generation is about two orders of magnitude smaller than $J_\mathrm{TE}$. The $J_\mathrm{QE}$ plot for $d = 500$ nm has a small dip near the flat-band voltage, at which  the effective tunneling distance (i.e., $x_2-x_1$) becomes the largest due to the flat potential profile. The maximum power density can be calculated by using the obtained $J_\mathrm{TE}$ and $J_\mathrm{QE}$ values, i.e., $P_\textrm{out} = [V( J_{\textrm{TE}}+J_{\textrm{QE}})]_{\textrm{max}}$. The solid square points marked in Fig. \ref{Fig3} denote the load voltages yielding the maximum power output, $V_\mathrm{max}$, which are located at around 0.6 V although the exact value depends on the gap distance. 

\begin{figure}[t!]
    \centering\includegraphics[width=1\linewidth]{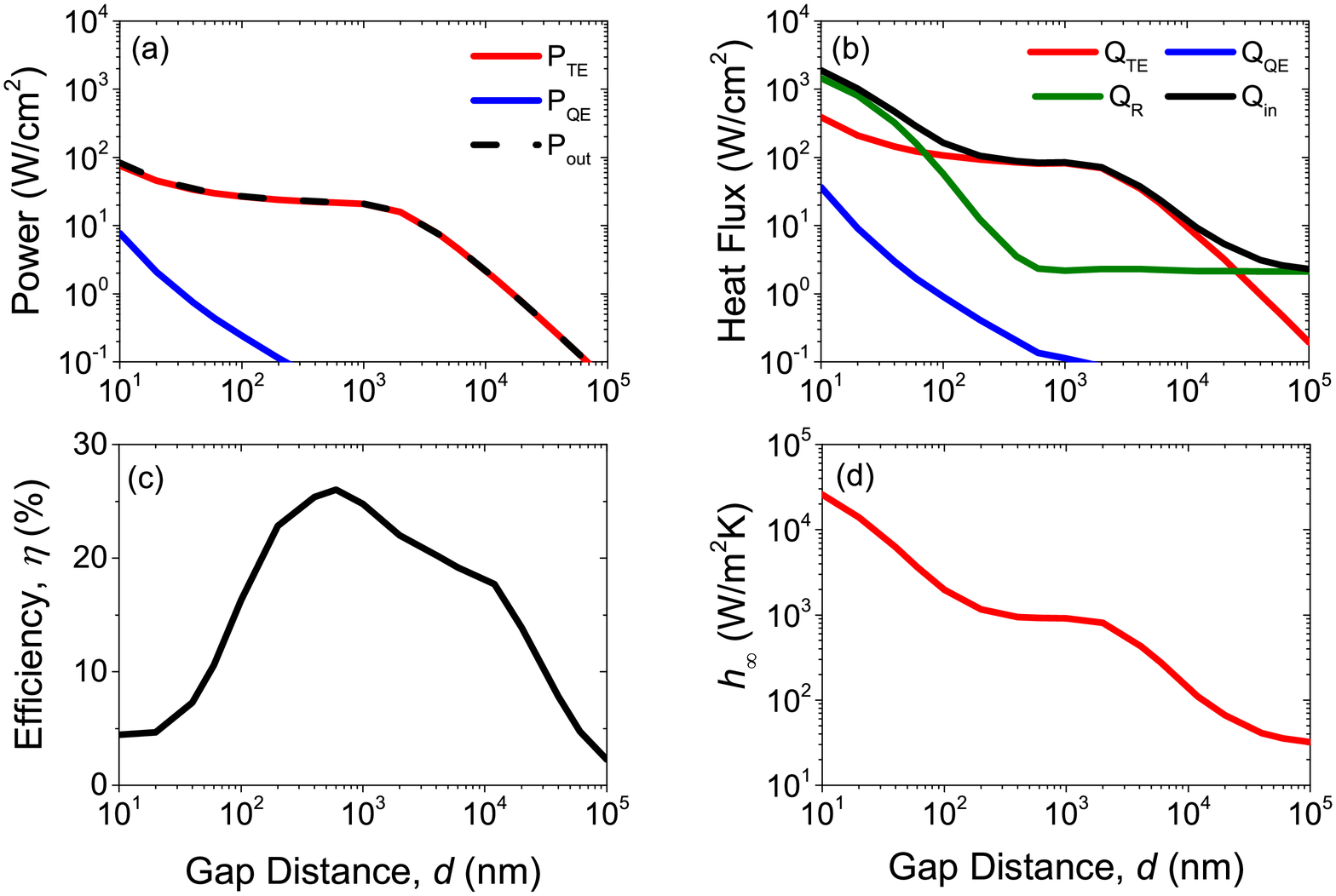}
    \caption{}
    \label{Fig4}
\end{figure}

Fig. \ref{Fig4} shows the gap-dependent performance of the TEC system when the emitter and collector temperatures are maintained at 1575 K ($\Phi_E=2.10$ eV) and 1000 K ($\Phi_C = 1.33$ eV), respectively. The net thermionic power density $P_\textrm{TE}$ increases by more than two orders of magnitude as the gap distance decreases, from less than 0.1 W/cm$^2$ at $d = 100$ $\mu$m to 22 W/cm$^2$ at $d = 500$ nm, indicating the eradication of space charges by the increased field strength across a small gap. The image charge effect gradually increases $P_\textrm{TE}$ as the vacuum gap decreases in the submicron regime, while more prominant enhancement is observed in sub-100 nm gap distances. The tunneling power density ($P_\mathrm{QE}$) also increases with the decreasing gap but contributes only up to $\sim$6\% of the total power output at $d=10$ nm, suggesting that the electron tunneling effect can be ignored unless an extremely small gap is to be considered. Conversely, Fig. \ref{Fig4}(b) shows the heat flux by electrons and thermal radiation. Both the thermionic heat flux ($Q_\textrm{TE}$) and the electron tunneling heat flux ($Q_\textrm{QE}$) follow similar trends to their respective power density curves in Fig. \ref{Fig4}(a). Therefore, thermionic emission is the dominant heat transfer mechanism in the vacuum gap range from $\sim$300 nm to $\sim$10 $\mu$m. On the other hand, the near-field enhancement of radiative heat transfer ($Q_\mathrm{R}$) significantly increases the total heat flux ($Q_\mathrm{in}$) for vacuum gaps smaller than 100 nm. In Fig. \ref{Fig4}(c), the energy conversion efficiency exhibits a maximum of $\sim$25 \% at $d \approx 500$ nm alongside a shoulder in the micrometer gap range. The obtained gap distance with the maximum efficiency is smaller than previously predicted (e.g., $0.9~ \mu \mathrm{m} \lesssim d \lesssim 3~\mu \mathrm{m}$ at $T_E = 1500$ K \cite{Lee2012j}). 
The convection heat transfer coefficient $h_\infty$ required to maintain the collector temperature at 1000 K is shown in Fig. \ref{Fig4}(d), suggesting that 300 nm $\lesssim d \lesssim$ 1 $\mu$m should be the optimal vacuum gap distance range for high thermionic efficiency while requiring a reasonable cooling load with $h_\infty$ in the order of 1000 W/m$^2$-K for thermal management. 

\begin{figure}[t!]
    \centering\includegraphics[width=1\linewidth]{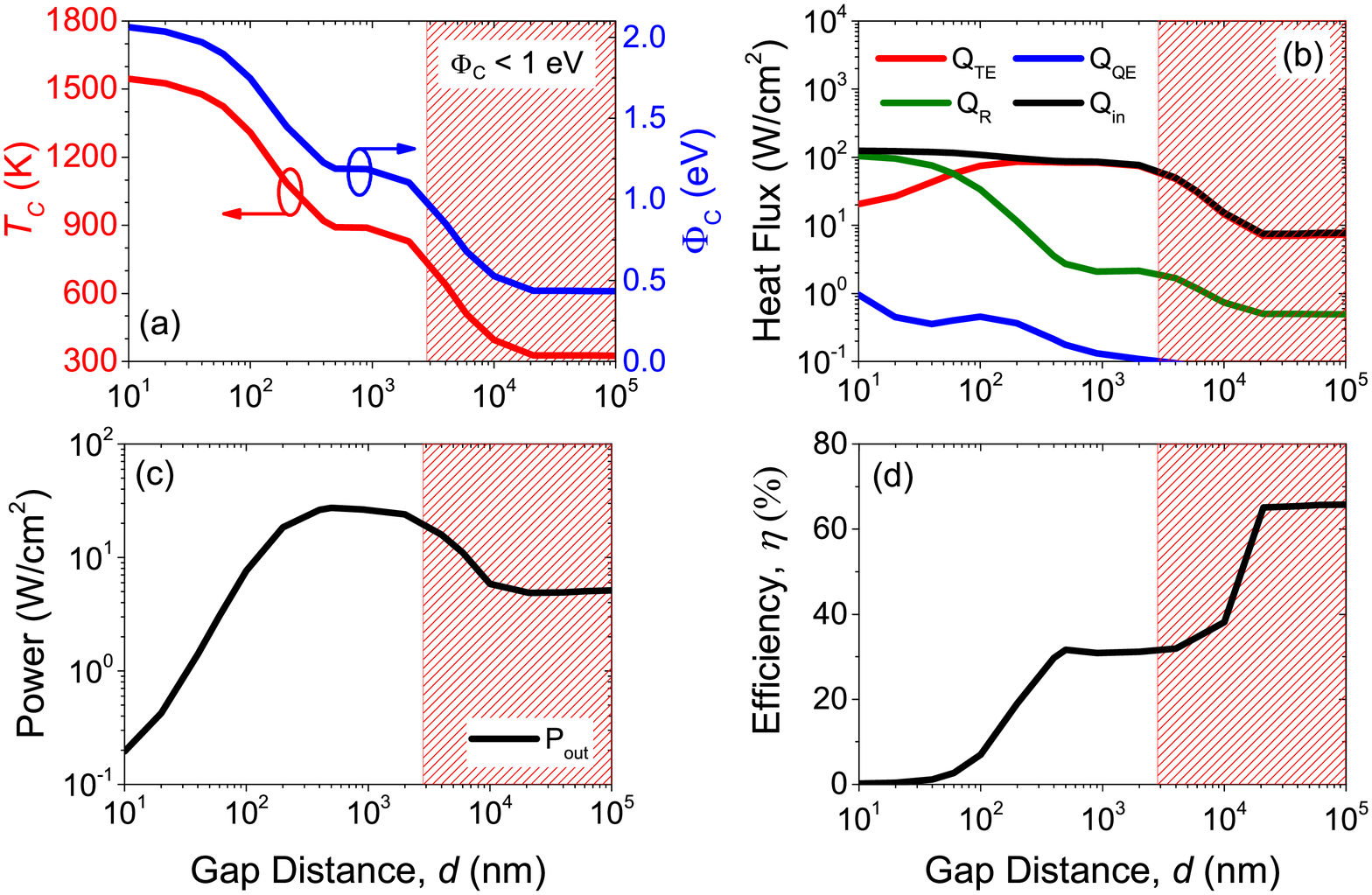}
    \caption{}
    \label{Fig5}
\end{figure}

In contrast to Fig. \ref{Fig4}, Fig. \ref{Fig5} shows TEC performances from a different design perspective by fixing the heat transfer coefficient at $h_\infty = 1000$ W/m$^2$-K. While the emitter temperature and work function remain the same as Fig. \ref{Fig4} (i.e., $T_E=1575$ K and $\Phi_E=2.10$ eV), $T_C$ (and subsequently $\Phi_C$ by $\Phi_C = T_C/750$) is adjusted to satisfy energy balance in the collector. For $d\gtrsim 10$ $\mu$m, the heat transfer coefficient is sufficient to cool the collector near room temperature, allowing sufficient thermionic charge transport even at such large gap distance while radiative heat transfer is limited by the far-field radiation. The consequent thermionic efficiency is higher than 60\%: see Fig. \ref{Fig5}(d) . However, since the lowest work function of a tungsten-barium-oxygen electrode has been predicted to be $\sim$1 eV \cite{Jacobs2017}, the obtained results in the red-hatched area may not be realistic. On the other hand, for small gaps ($d\lesssim 200$ nm), the collector temperature becomes nearly identical to the emitter temperature (e.g., $T_C=1546$ K and $\Phi_C=2.06$ eV at $d=10$ nm). Net thermionic power and heat flux thus significantly decrease due to large back emission from the collector, resulting in near zero thermionic efficiency. The optimal gap distance with the maximum power output (25 W/cm$^2$) and efficiency (32 \%) is determined to be $\sim$500 nm, at which $T_C = 890$ K and $\Phi_C = 1.19$ eV. Although the obtained optimum gap may change depending on desired operational design points, general trends discussed in Figs. \ref{Fig4} and \ref{Fig5} should be valid. Although the present study considers only one design condition at $T_E=1575$ K and $T_C=1000$ K, the emitter and collector temperature effects on thermionic power generation are further discussed in the Supplementary Information (Figs. S1 and S2). 

\begin{figure}[t!]
    \centering\includegraphics[width=0.8\linewidth]{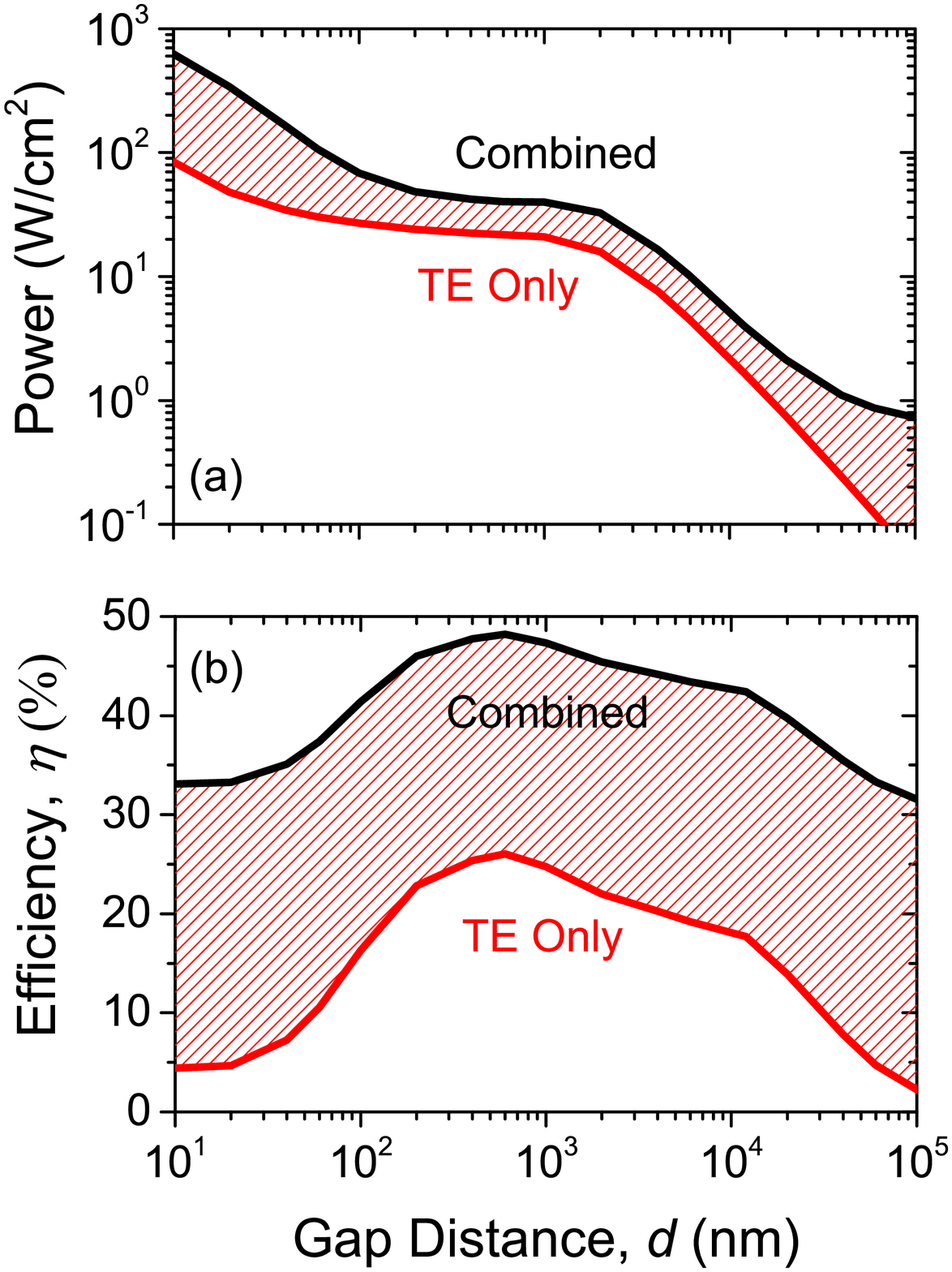}
    \caption{}
    \label{Fig6}
\end{figure}

While stand-alone submicron-gap TEC systems can achieve energy conversion efficiency at $\sim$30 \%, the thermionic collector can be used as a high-quality heat source for combined power generation \cite{Schwede2010}. Figure \ref{Fig6} shows the increase in power output and efficiency when the TEC device pairs with a conventional power system, such as a steam or Stirling heat engine \cite{Schwede2010,Xiao2017}, assumed to convert heat to electrical power at 30\% efficiency. The TEC system characteristics are $T_\textrm{E} = 1575$ K ($\Phi_\textrm{E} = 2.10$ eV) and $T_\textrm{C} = 1000$ K ($\Phi_\textrm{C} = 1.33$ eV). Both the power output and the efficiency exhibit significant enhancement when compared to the stand-alone TEC device, particularly for gaps less than 1 $\mu$m. The aforementioned near-field enhanced radiative heat transfer to the collector enables a significant amount of $Q_\textrm{out}$ delivered to the working fluid of the bottom cycle heat engine. This waste heat recycling almost doubles power generation from 22 W/cm$^2$ to 40 W/cm$^2$ at $d \approx 500$ nm, at which the combined cycle efficiency reaches a maximum of 48 \%. 

\begin{figure}[t!]
    \centering\includegraphics[width=0.9\linewidth]{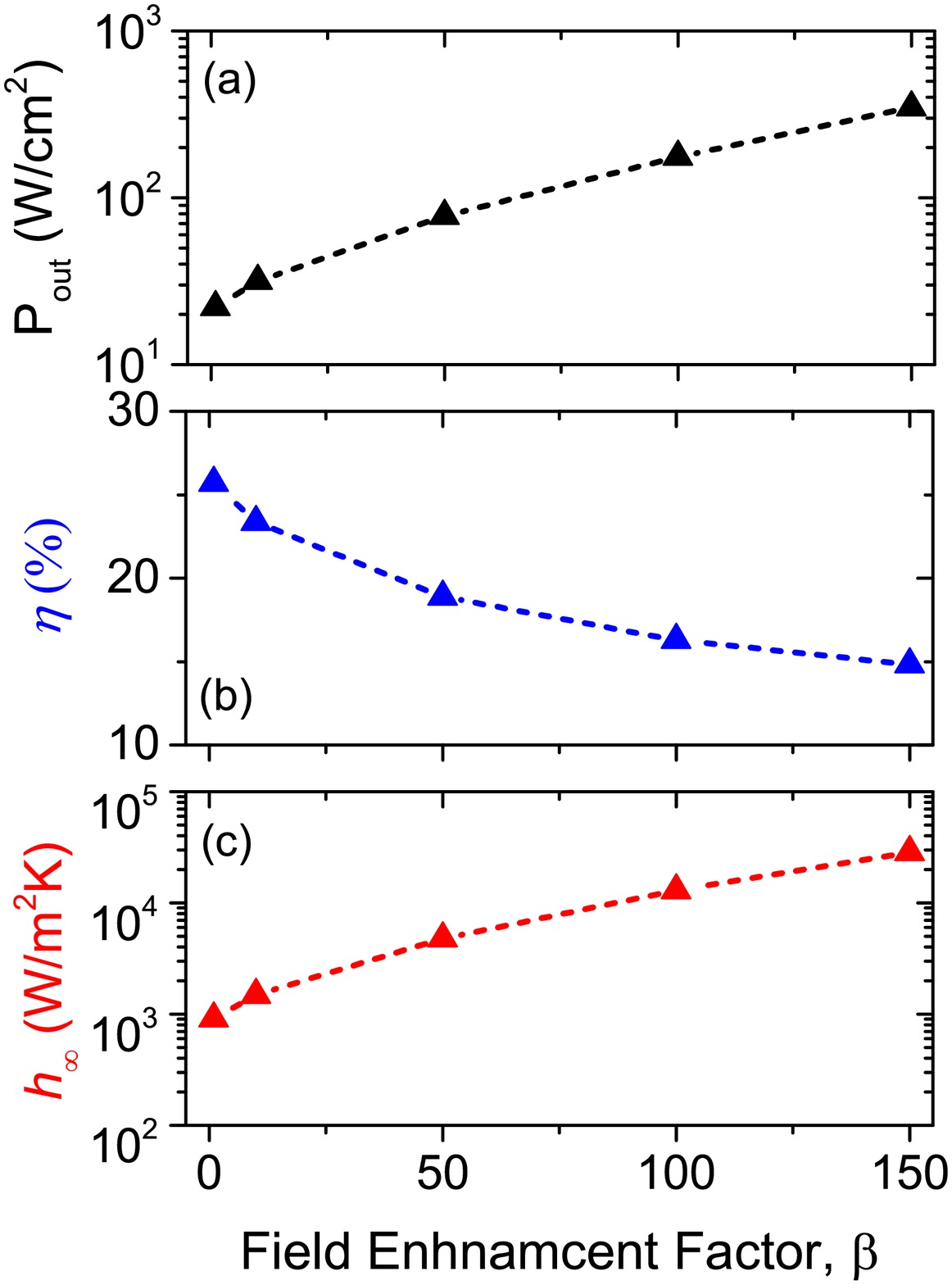}
    \caption{}
    \label{Fig7}
\end{figure}

Another approach to improve the submicron-gap TEC performance is to modify the potential barrier profile by locally concentrating the electric field on the electrode surfaces. Previous studies have found that local charge buildup at protrusion apexes of engineered surfaces having sharp tips or rough surface profiles enhances the local field near the emission site \cite{Miskovsky1993,Fisher2002c,Smith2006,Jensen2006}. When the surface peak apexes have a radius greater than 50 nm, the total concentrated field can be expressed as $E_c = \beta E$, where $\beta$ is a linear field enhancement factor \cite{Fisher2002c}. The potential barrier profile is then modified as \cite{Baldea2012a,Jensen2017b} 
\begin{equation}
    \label{Eqn:beta}
        W(x) = \Phi_E - \beta(\Phi_E - \Phi_C - qV)\left(\frac{x}{d}\right) +W_\textrm{ic}(x)
\end{equation}
to accommodate the field-induced potential profile change. Figure \ref{Fig7}(a) shows the impact of $\beta$ on $P_\textrm{out}$ at $d = 500$ nm for the same emitter and collector configurations as Fig. \ref{Fig4} (i.e., $T_E = 1575$ K and $T_C=1000$ K; $\Phi_E=2.1$ eV and $\Phi_C=1.3$ eV). As $\beta$ increases from 1 to 150, $P_\textrm{out}$ is enhanced by more than 15 times (i.e., 22 W/cm$^2$ to 348 W/cm$^2$), indicating that electrode surface engineering can greatly enhance thermionic emission of electrons. It should be noted that $\beta$ ranges from unity for a perfectly smooth surface to 140 for a typical dispenser cathode surface \cite{Jensen2006,Littau2013}. Field-induced electron emission also drastically enhances the thermionic heat transfer rate, which decreases the energy conversion efficiency as shown in Fig. \ref{Fig7}(b). For the calculation of the efficiency, we assume that near-field thermal radiation is not altered by rough surfaces. The heat transfer coefficient of a cooling fluid should increase to maintain the collector temperature as $\beta$ increases: see Fig. \ref{Fig7}(c), leading to the increase of heat loss to the environment. However, this heat loss can be recycled if the submicron-gap thermionic system is combined with a bottom-cycle heat engine. 

Experimental verification of submicron-gap thermionic power generation remains unexplored to date. Previous works have measured the effect of reducing a vacuum gap in thermionic energy conversion processes \cite{Littau2013,Lee2014,Yuan2017}. However, the minimum gap distance achieved to date is 11 $\mu$m by placing microspheres between electrodes as spacers \cite{Littau2013}. This limitation in the gap distance is mainly due to technical challenges in achieving submicron-gap distances between parallel planar structures. In addition, there was lack of evidences on the benefit of achieving such small gaps for thermionic power generation. \cite{Lee2012j}. However, recent progresses have been made to secure vacuum gaps on the order of 100 nm between mm$^2$-scale area plates, either by implementing microfabricated spacers \cite{Bernardi2016a,DeSutter2019} or nanopositioners \cite{Ghashami2018b}. The same strategy can be applied for the measurement of submicron-gap thermionic power generation, which will be a future research direction. 

\section{\label{sec:level1}Conclusions}
The present work has numerically studied thermionic energy conversion processes with an emphasis on comprehensive nanoscale charge and thermal transport phenomena. The obtained results demonstrate that submicron-gap TEC can generate a greater thermionic current density than the ideal case due to field-induced charge acceleration caused by strong electrostatic interactions with image charges. The systematic energy balance analysis yields the maximum energy conversion efficiency at around $25-30$\% when the device is operated at $T_E=1575$ K with the vacuum gap range of 300 nm $\lesssim d \lesssim$ 1 $\mu$m. This optimum gap range is smaller than previously determined in the micrometer scale. We also theoretically predicted potential advantages of implementing the TEC device into a combined power generation cycle and engineering the electrode surfaces to enhance the local field for amplified charge acceleration. Although the experimental demonstration of the predicted TEC performance is still future research, we believe that securing submicron vacuum gap distances will greatly enhance the performance of thermionic power generation, which will ultimately benefit direct thermal energy conversion technologies.

\bibliographystyle{apsrev4-1.bst}
\bibliography{References}

\begin{thebibliography}{47}%
\makeatletter
\providecommand \@ifxundefined [1]{%
 \@ifx{#1\undefined}
}%
\providecommand \@ifnum [1]{%
 \ifnum #1\expandafter \@firstoftwo
 \else \expandafter \@secondoftwo
 \fi
}%
\providecommand \@ifx [1]{%
 \ifx #1\expandafter \@firstoftwo
 \else \expandafter \@secondoftwo
 \fi
}%
\providecommand \natexlab [1]{#1}%
\providecommand \enquote  [1]{``#1''}%
\providecommand \bibnamefont  [1]{#1}%
\providecommand \bibfnamefont [1]{#1}%
\providecommand \citenamefont [1]{#1}%
\providecommand \href@noop [0]{\@secondoftwo}%
\providecommand \href [0]{\begingroup \@sanitize@url \@href}%
\providecommand \@href[1]{\@@startlink{#1}\@@href}%
\providecommand \@@href[1]{\endgroup#1\@@endlink}%
\providecommand \@sanitize@url [0]{\catcode `\\12\catcode `\$12\catcode
  `\&12\catcode `\#12\catcode `\^12\catcode `\_12\catcode `\%12\relax}%
\providecommand \@@startlink[1]{}%
\providecommand \@@endlink[0]{}%
\providecommand \url  [0]{\begingroup\@sanitize@url \@url }%
\providecommand \@url [1]{\endgroup\@href {#1}{\urlprefix }}%
\providecommand \urlprefix  [0]{URL }%
\providecommand \Eprint [0]{\href }%
\providecommand \doibase [0]{http://dx.doi.org/}%
\providecommand \selectlanguage [0]{\@gobble}%
\providecommand \bibinfo  [0]{\@secondoftwo}%
\providecommand \bibfield  [0]{\@secondoftwo}%
\providecommand \translation [1]{[#1]}%
\providecommand \BibitemOpen [0]{}%
\providecommand \bibitemStop [0]{}%
\providecommand \bibitemNoStop [0]{.\EOS\space}%
\providecommand \EOS [0]{\spacefactor3000\relax}%
\providecommand \BibitemShut  [1]{\csname bibitem#1\endcsname}%
\let\auto@bib@innerbib\@empty
\bibitem [{\citenamefont {Langmuir}(1923)}]{Langmuir1923}%
  \BibitemOpen
  \bibfield  {author} {\bibinfo {author} {\bibfnamefont {I.}~\bibnamefont
  {Langmuir}},\ }\href {\doibase 10.1103/PhysRev.21.419} {\bibfield  {journal}
  {\bibinfo  {journal} {Physical Review}\ }\textbf {\bibinfo {volume} {21}},\
  \bibinfo {pages} {419} (\bibinfo {year} {1923})}\BibitemShut {NoStop}%
\bibitem [{\citenamefont {Hatsopoulos}\ and\ \citenamefont
  {Gyftopoulos}(1973)}]{Hatsopoulos1973}%
  \BibitemOpen
  \bibfield  {author} {\bibinfo {author} {\bibfnamefont {N.}~\bibnamefont
  {Hatsopoulos}, \bibfnamefont {G.}}\ and\ \bibinfo {author} {\bibfnamefont
  {P.}~\bibnamefont {Gyftopoulos}, \bibfnamefont {E.}},\ }\href@noop {} {\emph
  {\bibinfo {title} {{Thermionic Energy Conversion}}}},\ Vol.~\bibinfo {volume}
  {1}\ (\bibinfo  {publisher} {MIT Press},\ \bibinfo {address} {Cambridge,
  Massachusetts},\ \bibinfo {year} {1973})\BibitemShut {NoStop}%
\bibitem [{\citenamefont {Lee}\ \emph {et~al.}(2014)\citenamefont {Lee},
  \citenamefont {Bargatin}, \citenamefont {Vancil}, \citenamefont {Gwinn},
  \citenamefont {Maboudian}, \citenamefont {Melosh},\ and\ \citenamefont
  {Howe}}]{Lee2014}%
  \BibitemOpen
  \bibfield  {author} {\bibinfo {author} {\bibfnamefont {J.~H.}\ \bibnamefont
  {Lee}}, \bibinfo {author} {\bibfnamefont {I.}~\bibnamefont {Bargatin}},
  \bibinfo {author} {\bibfnamefont {B.~K.}\ \bibnamefont {Vancil}}, \bibinfo
  {author} {\bibfnamefont {T.~O.}\ \bibnamefont {Gwinn}}, \bibinfo {author}
  {\bibfnamefont {R.}~\bibnamefont {Maboudian}}, \bibinfo {author}
  {\bibfnamefont {N.~A.}\ \bibnamefont {Melosh}}, \ and\ \bibinfo {author}
  {\bibfnamefont {R.~T.}\ \bibnamefont {Howe}},\ }\href {\doibase
  10.1109/JMEMS.2014.2307882} {\bibfield  {journal} {\bibinfo  {journal}
  {Journal of Microelectromechanical Systems}\ }\textbf {\bibinfo {volume}
  {23}},\ \bibinfo {pages} {1182} (\bibinfo {year} {2014})}\BibitemShut
  {NoStop}%
\bibitem [{\citenamefont {Shakouri}(2005)}]{Shakouri2005}%
  \BibitemOpen
  \bibfield  {author} {\bibinfo {author} {\bibfnamefont {A.}~\bibnamefont
  {Shakouri}},\ }in\ \href {\doibase 10.1109/ICT.2005.1519994} {\emph {\bibinfo
  {booktitle} {Proceedings of International Conference on Thermoelectronics}}}\
  (\bibinfo  {publisher} {IEEE},\ \bibinfo {address} {Clemson SC},\ \bibinfo
  {year} {2005})\ pp.\ \bibinfo {pages} {492--497}\BibitemShut {NoStop}%
\bibitem [{\citenamefont {Hishinuma}\ \emph {et~al.}(2001)\citenamefont
  {Hishinuma}, \citenamefont {Geballe}, \citenamefont {Moyzhes},\ and\
  \citenamefont {Kenny}}]{Hishinuma2001}%
  \BibitemOpen
  \bibfield  {author} {\bibinfo {author} {\bibfnamefont {Y.}~\bibnamefont
  {Hishinuma}}, \bibinfo {author} {\bibfnamefont {T.~H.}\ \bibnamefont
  {Geballe}}, \bibinfo {author} {\bibfnamefont {B.~Y.}\ \bibnamefont
  {Moyzhes}}, \ and\ \bibinfo {author} {\bibfnamefont {T.~W.}\ \bibnamefont
  {Kenny}},\ }\href {\doibase 10.1063/1.1365944} {\bibfield  {journal}
  {\bibinfo  {journal} {Applied Physics Letters}\ }\textbf {\bibinfo {volume}
  {78}},\ \bibinfo {pages} {2572} (\bibinfo {year} {2001})}\BibitemShut
  {NoStop}%
\bibitem [{\citenamefont {Skoplaki}\ and\ \citenamefont
  {Palyvos}(2009)}]{Skoplaki2009}%
  \BibitemOpen
  \bibfield  {author} {\bibinfo {author} {\bibfnamefont {E.}~\bibnamefont
  {Skoplaki}}\ and\ \bibinfo {author} {\bibfnamefont {J.~A.}\ \bibnamefont
  {Palyvos}},\ }\href {\doibase 10.1016/j.solener.2008.10.008} {\bibfield
  {journal} {\bibinfo  {journal} {Solar Energy}\ }\textbf {\bibinfo {volume}
  {83}},\ \bibinfo {pages} {614} (\bibinfo {year} {2009})}\BibitemShut
  {NoStop}%
\bibitem [{\citenamefont {Zhu}\ \emph {et~al.}(2015)\citenamefont {Zhu},
  \citenamefont {Raman},\ and\ \citenamefont {Fan}}]{Zhu2015}%
  \BibitemOpen
  \bibfield  {author} {\bibinfo {author} {\bibfnamefont {L.}~\bibnamefont
  {Zhu}}, \bibinfo {author} {\bibfnamefont {A.~P.}\ \bibnamefont {Raman}}, \
  and\ \bibinfo {author} {\bibfnamefont {S.}~\bibnamefont {Fan}},\ }\href
  {\doibase 10.1073/pnas.1509453112} {\bibfield  {journal} {\bibinfo  {journal}
  {Proceedings of the National Academy of Sciences}\ }\textbf {\bibinfo
  {volume} {112}},\ \bibinfo {pages} {12282} (\bibinfo {year}
  {2015})}\BibitemShut {NoStop}%
\bibitem [{\citenamefont {Yuan}\ \emph {et~al.}(2017)\citenamefont {Yuan},
  \citenamefont {Riley}, \citenamefont {Shen}, \citenamefont {Pianetta},
  \citenamefont {Melosh},\ and\ \citenamefont {Howe}}]{Yuan2017}%
  \BibitemOpen
  \bibfield  {author} {\bibinfo {author} {\bibfnamefont {H.}~\bibnamefont
  {Yuan}}, \bibinfo {author} {\bibfnamefont {D.~C.}\ \bibnamefont {Riley}},
  \bibinfo {author} {\bibfnamefont {Z.-X.}\ \bibnamefont {Shen}}, \bibinfo
  {author} {\bibfnamefont {P.~A.}\ \bibnamefont {Pianetta}}, \bibinfo {author}
  {\bibfnamefont {N.~A.}\ \bibnamefont {Melosh}}, \ and\ \bibinfo {author}
  {\bibfnamefont {R.~T.}\ \bibnamefont {Howe}},\ }\href {\doibase
  10.1016/j.nanoen.2016.12.027} {\bibfield  {journal} {\bibinfo  {journal}
  {Nano Energy}\ }\textbf {\bibinfo {volume} {32}},\ \bibinfo {pages} {67}
  (\bibinfo {year} {2017})}\BibitemShut {NoStop}%
\bibitem [{\citenamefont {Koeck}\ and\ \citenamefont
  {Nemanich}(2012)}]{Koeck2012}%
  \BibitemOpen
  \bibfield  {author} {\bibinfo {author} {\bibfnamefont {F.~A.}\ \bibnamefont
  {Koeck}}\ and\ \bibinfo {author} {\bibfnamefont {R.~J.}\ \bibnamefont
  {Nemanich}},\ }\href {\doibase 10.1063/1.4766442} {\bibfield  {journal}
  {\bibinfo  {journal} {Journal of Applied Physics}\ }\textbf {\bibinfo
  {volume} {112}},\ \bibinfo {pages} {113707} (\bibinfo {year}
  {2012})}\BibitemShut {NoStop}%
\bibitem [{\citenamefont {Littau}\ \emph {et~al.}(2013)\citenamefont {Littau},
  \citenamefont {Sahasrabuddhe}, \citenamefont {Barfield}, \citenamefont
  {Yuan}, \citenamefont {Shen}, \citenamefont {Howe},\ and\ \citenamefont
  {Melosh}}]{Littau2013}%
  \BibitemOpen
  \bibfield  {author} {\bibinfo {author} {\bibfnamefont {K.~A.}\ \bibnamefont
  {Littau}}, \bibinfo {author} {\bibfnamefont {K.}~\bibnamefont
  {Sahasrabuddhe}}, \bibinfo {author} {\bibfnamefont {D.}~\bibnamefont
  {Barfield}}, \bibinfo {author} {\bibfnamefont {H.}~\bibnamefont {Yuan}},
  \bibinfo {author} {\bibfnamefont {Z.-X.}\ \bibnamefont {Shen}}, \bibinfo
  {author} {\bibfnamefont {R.~T.}\ \bibnamefont {Howe}}, \ and\ \bibinfo
  {author} {\bibfnamefont {N.~A.}\ \bibnamefont {Melosh}},\ }\href {\doibase
  10.1039/c3cp52895b} {\bibfield  {journal} {\bibinfo  {journal} {Physical
  Chemistry Chemical Physics}\ }\textbf {\bibinfo {volume} {15}},\ \bibinfo
  {pages} {14442} (\bibinfo {year} {2013})}\BibitemShut {NoStop}%
\bibitem [{\citenamefont {Suzuki}\ \emph {et~al.}(2009)\citenamefont {Suzuki},
  \citenamefont {Ono}, \citenamefont {Sakuma},\ and\ \citenamefont
  {Sakai}}]{Suzuki2009}%
  \BibitemOpen
  \bibfield  {author} {\bibinfo {author} {\bibfnamefont {M.}~\bibnamefont
  {Suzuki}}, \bibinfo {author} {\bibfnamefont {T.}~\bibnamefont {Ono}},
  \bibinfo {author} {\bibfnamefont {N.}~\bibnamefont {Sakuma}}, \ and\ \bibinfo
  {author} {\bibfnamefont {T.}~\bibnamefont {Sakai}},\ }\href {\doibase
  10.1016/j.diamond.2009.05.004} {\bibfield  {journal} {\bibinfo  {journal}
  {Diamond and Related Materials}\ }\textbf {\bibinfo {volume} {18}},\ \bibinfo
  {pages} {1274} (\bibinfo {year} {2009})}\BibitemShut {NoStop}%
\bibitem [{\citenamefont {Voss}\ \emph {et~al.}(2014)\citenamefont {Voss},
  \citenamefont {Vojvodic}, \citenamefont {Chou}, \citenamefont {Howe},\ and\
  \citenamefont {Abild-Pedersen}}]{Voss2014}%
  \BibitemOpen
  \bibfield  {author} {\bibinfo {author} {\bibfnamefont {J.}~\bibnamefont
  {Voss}}, \bibinfo {author} {\bibfnamefont {A.}~\bibnamefont {Vojvodic}},
  \bibinfo {author} {\bibfnamefont {S.~H.}\ \bibnamefont {Chou}}, \bibinfo
  {author} {\bibfnamefont {R.~T.}\ \bibnamefont {Howe}}, \ and\ \bibinfo
  {author} {\bibfnamefont {F.}~\bibnamefont {Abild-Pedersen}},\ }\href
  {\doibase 10.1103/PhysRevApplied.2.024004} {\bibfield  {journal} {\bibinfo
  {journal} {Physical Review Applied}\ }\textbf {\bibinfo {volume} {2}},\
  \bibinfo {pages} {1} (\bibinfo {year} {2014})}\BibitemShut {NoStop}%
\bibitem [{\citenamefont {Kato}\ \emph {et~al.}(2016)\citenamefont {Kato},
  \citenamefont {Takeuchi}, \citenamefont {Ogura}, \citenamefont {Yamada},
  \citenamefont {Kataoka}, \citenamefont {Kimura}, \citenamefont {Sobue},
  \citenamefont {Nebel},\ and\ \citenamefont {Yamasaki}}]{Kato2016}%
  \BibitemOpen
  \bibfield  {author} {\bibinfo {author} {\bibfnamefont {H.}~\bibnamefont
  {Kato}}, \bibinfo {author} {\bibfnamefont {D.}~\bibnamefont {Takeuchi}},
  \bibinfo {author} {\bibfnamefont {M.}~\bibnamefont {Ogura}}, \bibinfo
  {author} {\bibfnamefont {T.}~\bibnamefont {Yamada}}, \bibinfo {author}
  {\bibfnamefont {M.}~\bibnamefont {Kataoka}}, \bibinfo {author} {\bibfnamefont
  {Y.}~\bibnamefont {Kimura}}, \bibinfo {author} {\bibfnamefont
  {S.}~\bibnamefont {Sobue}}, \bibinfo {author} {\bibfnamefont {C.~E.}\
  \bibnamefont {Nebel}}, \ and\ \bibinfo {author} {\bibfnamefont
  {S.}~\bibnamefont {Yamasaki}},\ }\href {\doibase
  10.1016/j.diamond.2015.08.002} {\bibfield  {journal} {\bibinfo  {journal}
  {Diamond and Related Materials}\ }\textbf {\bibinfo {volume} {63}},\ \bibinfo
  {pages} {165} (\bibinfo {year} {2016})}\BibitemShut {NoStop}%
\bibitem [{\citenamefont {Barmina}\ \emph {et~al.}(2012)\citenamefont
  {Barmina}, \citenamefont {Serkov}, \citenamefont {Stratakis}, \citenamefont
  {Fotakis}, \citenamefont {Stolyarov}, \citenamefont {Stolyarov},\ and\
  \citenamefont {Shafeev}}]{Barmina2012}%
  \BibitemOpen
  \bibfield  {author} {\bibinfo {author} {\bibfnamefont {E.~V.}\ \bibnamefont
  {Barmina}}, \bibinfo {author} {\bibfnamefont {A.~A.}\ \bibnamefont {Serkov}},
  \bibinfo {author} {\bibfnamefont {E.}~\bibnamefont {Stratakis}}, \bibinfo
  {author} {\bibfnamefont {C.}~\bibnamefont {Fotakis}}, \bibinfo {author}
  {\bibfnamefont {V.~N.}\ \bibnamefont {Stolyarov}}, \bibinfo {author}
  {\bibfnamefont {I.~N.}\ \bibnamefont {Stolyarov}}, \ and\ \bibinfo {author}
  {\bibfnamefont {G.~A.}\ \bibnamefont {Shafeev}},\ }\href {\doibase
  10.1007/s00339-011-6692-6} {\bibfield  {journal} {\bibinfo  {journal}
  {Applied Physics A: Materials Science and Processing}\ }\textbf {\bibinfo
  {volume} {106}},\ \bibinfo {pages} {1} (\bibinfo {year} {2012})}\BibitemShut
  {NoStop}%
\bibitem [{\citenamefont {Hatsopoulos}\ and\ \citenamefont
  {Gyftopoulos}(1979)}]{Hatsopoulos1979}%
  \BibitemOpen
  \bibfield  {author} {\bibinfo {author} {\bibfnamefont {N.}~\bibnamefont
  {Hatsopoulos}, \bibfnamefont {G.}}\ and\ \bibinfo {author} {\bibfnamefont
  {P.}~\bibnamefont {Gyftopoulos}, \bibfnamefont {E.}},\ }\href@noop {} {\emph
  {\bibinfo {title} {{Thermionic Energy Conversion}}}},\ Vol.~\bibinfo {volume}
  {2}\ (\bibinfo  {publisher} {MIT Press},\ \bibinfo {address} {Cambridge,
  Massachusetts},\ \bibinfo {year} {1979})\BibitemShut {NoStop}%
\bibitem [{\citenamefont {Lee}\ \emph {et~al.}(2012)\citenamefont {Lee},
  \citenamefont {Bargatin}, \citenamefont {Melosh},\ and\ \citenamefont
  {Howe}}]{Lee2012j}%
  \BibitemOpen
  \bibfield  {author} {\bibinfo {author} {\bibfnamefont {J.~H.}\ \bibnamefont
  {Lee}}, \bibinfo {author} {\bibfnamefont {I.}~\bibnamefont {Bargatin}},
  \bibinfo {author} {\bibfnamefont {N.~A.}\ \bibnamefont {Melosh}}, \ and\
  \bibinfo {author} {\bibfnamefont {R.~T.}\ \bibnamefont {Howe}},\ }\href
  {\doibase 10.1063/1.4707379} {\bibfield  {journal} {\bibinfo  {journal}
  {Applied Physics Letters}\ }\textbf {\bibinfo {volume} {100}},\ \bibinfo
  {pages} {173904} (\bibinfo {year} {2012})}\BibitemShut {NoStop}%
\bibitem [{\citenamefont {Jensen}(2017)}]{Jensen2017b}%
  \BibitemOpen
  \bibfield  {author} {\bibinfo {author} {\bibfnamefont {K.~L.}\ \bibnamefont
  {Jensen}},\ }\href {\doibase 10.1002/9781119051794.part2} {\emph {\bibinfo
  {title} {{Introduction to the Physics of Electron Emission: Theory and
  Simulation}}}}\ (\bibinfo  {publisher} {John Wiley {\&} Sons},\ \bibinfo
  {address} {Chichester, UK},\ \bibinfo {year} {2017})\BibitemShut {NoStop}%
\bibitem [{\citenamefont {Lim}\ \emph {et~al.}(2018)\citenamefont {Lim},
  \citenamefont {Lambert}, \citenamefont {Vay},\ and\ \citenamefont
  {Schwede}}]{Lim2018}%
  \BibitemOpen
  \bibfield  {author} {\bibinfo {author} {\bibfnamefont {I.~T.}\ \bibnamefont
  {Lim}}, \bibinfo {author} {\bibfnamefont {S.~A.}\ \bibnamefont {Lambert}},
  \bibinfo {author} {\bibfnamefont {J.~L.}\ \bibnamefont {Vay}}, \ and\
  \bibinfo {author} {\bibfnamefont {J.~W.}\ \bibnamefont {Schwede}},\ }\href
  {\doibase 10.1063/1.5018067} {\bibfield  {journal} {\bibinfo  {journal}
  {Applied Physics Letters}\ }\textbf {\bibinfo {volume} {112}},\ \bibinfo
  {pages} {073906} (\bibinfo {year} {2018})}\BibitemShut {NoStop}%
\bibitem [{\citenamefont {Xiao}\ \emph {et~al.}(2017)\citenamefont {Xiao},
  \citenamefont {Zheng}, \citenamefont {Qiu}, \citenamefont {Li}, \citenamefont
  {Li},\ and\ \citenamefont {Ni}}]{Xiao2017}%
  \BibitemOpen
  \bibfield  {author} {\bibinfo {author} {\bibfnamefont {G.}~\bibnamefont
  {Xiao}}, \bibinfo {author} {\bibfnamefont {G.}~\bibnamefont {Zheng}},
  \bibinfo {author} {\bibfnamefont {M.}~\bibnamefont {Qiu}}, \bibinfo {author}
  {\bibfnamefont {Q.}~\bibnamefont {Li}}, \bibinfo {author} {\bibfnamefont
  {D.}~\bibnamefont {Li}}, \ and\ \bibinfo {author} {\bibfnamefont
  {M.}~\bibnamefont {Ni}},\ }\href {\doibase 10.1016/j.apenergy.2017.09.021}
  {\bibfield  {journal} {\bibinfo  {journal} {Applied Energy}\ }\textbf
  {\bibinfo {volume} {208}},\ \bibinfo {pages} {1318} (\bibinfo {year}
  {2017})}\BibitemShut {NoStop}%
\bibitem [{\citenamefont {Meir}\ \emph {et~al.}(2013)\citenamefont {Meir},
  \citenamefont {Stephanos}, \citenamefont {Geballe},\ and\ \citenamefont
  {Mannhart}}]{Meir2013a}%
  \BibitemOpen
  \bibfield  {author} {\bibinfo {author} {\bibfnamefont {S.}~\bibnamefont
  {Meir}}, \bibinfo {author} {\bibfnamefont {C.}~\bibnamefont {Stephanos}},
  \bibinfo {author} {\bibfnamefont {T.~H.}\ \bibnamefont {Geballe}}, \ and\
  \bibinfo {author} {\bibfnamefont {J.}~\bibnamefont {Mannhart}},\ }\href
  {\doibase 10.1063/1.4817730} {\bibfield  {journal} {\bibinfo  {journal}
  {Journal of Renewable and Sustainable Energy}\ }\textbf {\bibinfo {volume}
  {5}},\ \bibinfo {pages} {043127} (\bibinfo {year} {2013})}\BibitemShut
  {NoStop}%
\bibitem [{\citenamefont {Wanke}\ \emph {et~al.}(2016)\citenamefont {Wanke},
  \citenamefont {Hassink}, \citenamefont {Stephanos}, \citenamefont {Rastegar},
  \citenamefont {Braun},\ and\ \citenamefont {Mannhart}}]{Wanke2016}%
  \BibitemOpen
  \bibfield  {author} {\bibinfo {author} {\bibfnamefont {R.}~\bibnamefont
  {Wanke}}, \bibinfo {author} {\bibfnamefont {G.~W.}\ \bibnamefont {Hassink}},
  \bibinfo {author} {\bibfnamefont {C.}~\bibnamefont {Stephanos}}, \bibinfo
  {author} {\bibfnamefont {I.}~\bibnamefont {Rastegar}}, \bibinfo {author}
  {\bibfnamefont {W.}~\bibnamefont {Braun}}, \ and\ \bibinfo {author}
  {\bibfnamefont {J.}~\bibnamefont {Mannhart}},\ }\href {\doibase
  10.1063/1.4955073} {\bibfield  {journal} {\bibinfo  {journal} {Journal of
  Applied Physics}\ }\textbf {\bibinfo {volume} {119}} (\bibinfo {year}
  {2016}),\ 10.1063/1.4955073}\BibitemShut {NoStop}%
\bibitem [{\citenamefont {Zeng}(2006)}]{Zeng2006}%
  \BibitemOpen
  \bibfield  {author} {\bibinfo {author} {\bibfnamefont {T.}~\bibnamefont
  {Zeng}},\ }\href {\doibase 10.1063/1.2192973} {\bibfield  {journal} {\bibinfo
   {journal} {Applied Physics Letters}\ }\textbf {\bibinfo {volume} {88}},\
  \bibinfo {pages} {153104} (\bibinfo {year} {2006})}\BibitemShut {NoStop}%
\bibitem [{\citenamefont {O'Dwyer}\ \emph {et~al.}(2009)\citenamefont
  {O'Dwyer}, \citenamefont {Humphrey}, \citenamefont {Lewis},\ and\
  \citenamefont {Zhang}}]{ODwyer2009a}%
  \BibitemOpen
  \bibfield  {author} {\bibinfo {author} {\bibfnamefont {M.~F.}\ \bibnamefont
  {O'Dwyer}}, \bibinfo {author} {\bibfnamefont {T.~E.}\ \bibnamefont
  {Humphrey}}, \bibinfo {author} {\bibfnamefont {R.~A.}\ \bibnamefont {Lewis}},
  \ and\ \bibinfo {author} {\bibfnamefont {C.}~\bibnamefont {Zhang}},\ }\href
  {\doibase 10.1088/0022-3727/42/3/035417} {\bibfield  {journal} {\bibinfo
  {journal} {Journal of Physics D}\ }\textbf {\bibinfo {volume} {42}},\
  \bibinfo {pages} {035417} (\bibinfo {year} {2009})}\BibitemShut {NoStop}%
\bibitem [{\citenamefont {Lee}\ \emph {et~al.}(2009)\citenamefont {Lee},
  \citenamefont {Jeong}, \citenamefont {No}, \citenamefont {Hannebauer},\ and\
  \citenamefont {Yoo}}]{Lee2009f}%
  \BibitemOpen
  \bibfield  {author} {\bibinfo {author} {\bibfnamefont {J.~I.}\ \bibnamefont
  {Lee}}, \bibinfo {author} {\bibfnamefont {Y.~H.}\ \bibnamefont {Jeong}},
  \bibinfo {author} {\bibfnamefont {H.~C.}\ \bibnamefont {No}}, \bibinfo
  {author} {\bibfnamefont {R.}~\bibnamefont {Hannebauer}}, \ and\ \bibinfo
  {author} {\bibfnamefont {S.~K.}\ \bibnamefont {Yoo}},\ }\href {\doibase
  10.1063/1.3266921} {\bibfield  {journal} {\bibinfo  {journal} {Applied
  Physics Letters}\ }\textbf {\bibinfo {volume} {95}},\ \bibinfo {pages}
  {223107} (\bibinfo {year} {2009})}\BibitemShut {NoStop}%
\bibitem [{\citenamefont {Voss}\ \emph {et~al.}(2013)\citenamefont {Voss},
  \citenamefont {Vojvodic}, \citenamefont {Chou}, \citenamefont {Howe},
  \citenamefont {Bargatin},\ and\ \citenamefont {Abild-Pedersen}}]{Voss2013}%
  \BibitemOpen
  \bibfield  {author} {\bibinfo {author} {\bibfnamefont {J.}~\bibnamefont
  {Voss}}, \bibinfo {author} {\bibfnamefont {A.}~\bibnamefont {Vojvodic}},
  \bibinfo {author} {\bibfnamefont {S.~H.}\ \bibnamefont {Chou}}, \bibinfo
  {author} {\bibfnamefont {R.~T.}\ \bibnamefont {Howe}}, \bibinfo {author}
  {\bibfnamefont {I.}~\bibnamefont {Bargatin}}, \ and\ \bibinfo {author}
  {\bibfnamefont {F.}~\bibnamefont {Abild-Pedersen}},\ }\href {\doibase
  10.1063/1.4805002} {\bibfield  {journal} {\bibinfo  {journal} {Journal of
  Chemical Physics}\ }\textbf {\bibinfo {volume} {138}},\ \bibinfo {pages}
  {204701} (\bibinfo {year} {2013})}\BibitemShut {NoStop}%
\bibitem [{\citenamefont {Wang}\ \emph {et~al.}(2016)\citenamefont {Wang},
  \citenamefont {Liao}, \citenamefont {Zhang}, \citenamefont {Chen},
  \citenamefont {Su},\ and\ \citenamefont {Chen}}]{Wang2016}%
  \BibitemOpen
  \bibfield  {author} {\bibinfo {author} {\bibfnamefont {Y.}~\bibnamefont
  {Wang}}, \bibinfo {author} {\bibfnamefont {T.}~\bibnamefont {Liao}}, \bibinfo
  {author} {\bibfnamefont {Y.}~\bibnamefont {Zhang}}, \bibinfo {author}
  {\bibfnamefont {X.}~\bibnamefont {Chen}}, \bibinfo {author} {\bibfnamefont
  {S.}~\bibnamefont {Su}}, \ and\ \bibinfo {author} {\bibfnamefont
  {J.}~\bibnamefont {Chen}},\ }\href {\doibase 10.1063/1.4940720} {\bibfield
  {journal} {\bibinfo  {journal} {Journal of Applied Physics}\ }\textbf
  {\bibinfo {volume} {119}},\ \bibinfo {pages} {045106} (\bibinfo {year}
  {2016})}\BibitemShut {NoStop}%
\bibitem [{\citenamefont {O'Dwyer}\ \emph {et~al.}(2005)\citenamefont
  {O'Dwyer}, \citenamefont {Lewis}, \citenamefont {Zhang},\ and\ \citenamefont
  {Humphrey}}]{ODwyer2005a}%
  \BibitemOpen
  \bibfield  {author} {\bibinfo {author} {\bibfnamefont {M.~F.}\ \bibnamefont
  {O'Dwyer}}, \bibinfo {author} {\bibfnamefont {R.~A.}\ \bibnamefont {Lewis}},
  \bibinfo {author} {\bibfnamefont {C.}~\bibnamefont {Zhang}}, \ and\ \bibinfo
  {author} {\bibfnamefont {T.~E.}\ \bibnamefont {Humphrey}},\ }\href {\doibase
  10.1103/PhysRevB.72.205330} {\bibfield  {journal} {\bibinfo  {journal}
  {Physical Review B}\ }\textbf {\bibinfo {volume} {72}},\ \bibinfo {pages}
  {205330} (\bibinfo {year} {2005})}\BibitemShut {NoStop}%
\bibitem [{\citenamefont {Murphy}\ and\ \citenamefont
  {Good}(1956)}]{Murphy1956}%
  \BibitemOpen
  \bibfield  {author} {\bibinfo {author} {\bibfnamefont {E.~L.}\ \bibnamefont
  {Murphy}}\ and\ \bibinfo {author} {\bibfnamefont {R.~H.}\ \bibnamefont
  {Good}},\ }\href {\doibase 10.1103/PhysRev.102.1464} {\bibfield  {journal}
  {\bibinfo  {journal} {Physical Review}\ }\textbf {\bibinfo {volume} {102}},\
  \bibinfo {pages} {1464} (\bibinfo {year} {1956})}\BibitemShut {NoStop}%
\bibitem [{\citenamefont {Christov}(1966)}]{Christov1966a}%
  \BibitemOpen
  \bibfield  {author} {\bibinfo {author} {\bibfnamefont {S.~G.}\ \bibnamefont
  {Christov}},\ }\href@noop {} {\bibfield  {journal} {\bibinfo  {journal}
  {Physica Status Solidi b}\ }\textbf {\bibinfo {volume} {17}},\ \bibinfo
  {pages} {11} (\bibinfo {year} {1966})}\BibitemShut {NoStop}%
\bibitem [{\citenamefont {Jensen}(2003)}]{Jensen2003}%
  \BibitemOpen
  \bibfield  {author} {\bibinfo {author} {\bibfnamefont {K.~L.}\ \bibnamefont
  {Jensen}},\ }\href {\doibase 10.1116/1.1573664} {\bibfield  {journal}
  {\bibinfo  {journal} {Journal of Vacuum Science {\&} Technology B}\ }\textbf
  {\bibinfo {volume} {21}},\ \bibinfo {pages} {1528} (\bibinfo {year}
  {2003})}\BibitemShut {NoStop}%
\bibitem [{\citenamefont {Baeva}(2018)}]{Baeva2018}%
  \BibitemOpen
  \bibfield  {author} {\bibinfo {author} {\bibfnamefont {M.}~\bibnamefont
  {Baeva}},\ }\href {\doibase 10.1063/1.5041314} {\bibfield  {journal}
  {\bibinfo  {journal} {AIP Advances}\ }\textbf {\bibinfo {volume} {8}},\
  \bibinfo {pages} {85322} (\bibinfo {year} {2018})}\BibitemShut {NoStop}%
\bibitem [{\citenamefont {Simmons}(1963)}]{Simmons1963a}%
  \BibitemOpen
  \bibfield  {author} {\bibinfo {author} {\bibfnamefont {J.~G.}\ \bibnamefont
  {Simmons}},\ }\href {\doibase 10.1063/1.1735973} {\bibfield  {journal}
  {\bibinfo  {journal} {Journal of Applied Physics}\ }\textbf {\bibinfo
  {volume} {34}},\ \bibinfo {pages} {2581} (\bibinfo {year}
  {1963})}\BibitemShut {NoStop}%
\bibitem [{\citenamefont {Baldea}\ and\ \citenamefont
  {Koppel}(2012)}]{Baldea2012a}%
  \BibitemOpen
  \bibfield  {author} {\bibinfo {author} {\bibfnamefont {I.}~\bibnamefont
  {Baldea}}\ and\ \bibinfo {author} {\bibfnamefont {H.}~\bibnamefont
  {Koppel}},\ }\href {\doibase 10.1002/pssb.201248034} {\bibfield  {journal}
  {\bibinfo  {journal} {Physica Status Solidi B}\ }\textbf {\bibinfo {volume}
  {249}},\ \bibinfo {pages} {1791} (\bibinfo {year} {2012})}\BibitemShut
  {NoStop}%
\bibitem [{\citenamefont {Park}\ and\ \citenamefont {Zhang}(2013)}]{Park2013}%
  \BibitemOpen
  \bibfield  {author} {\bibinfo {author} {\bibfnamefont {K.}~\bibnamefont
  {Park}}\ and\ \bibinfo {author} {\bibfnamefont {Z.~M.}\ \bibnamefont
  {Zhang}},\ }\href {\doibase 10.5098/hmt.v4.1.3001} {\bibfield  {journal}
  {\bibinfo  {journal} {Frontiers in Heat and Mass Transfer}\ }\textbf
  {\bibinfo {volume} {4}},\ \bibinfo {pages} {013001} (\bibinfo {year}
  {2013})}\BibitemShut {NoStop}%
\bibitem [{\citenamefont {Joulain}\ \emph {et~al.}(2005)\citenamefont
  {Joulain}, \citenamefont {Mulet}, \citenamefont {Marquier}, \citenamefont
  {Carminati},\ and\ \citenamefont {Greffet}}]{Joulain2005c}%
  \BibitemOpen
  \bibfield  {author} {\bibinfo {author} {\bibfnamefont {K.}~\bibnamefont
  {Joulain}}, \bibinfo {author} {\bibfnamefont {J.-P.}\ \bibnamefont {Mulet}},
  \bibinfo {author} {\bibfnamefont {F.}~\bibnamefont {Marquier}}, \bibinfo
  {author} {\bibfnamefont {R.}~\bibnamefont {Carminati}}, \ and\ \bibinfo
  {author} {\bibfnamefont {J.-J.}\ \bibnamefont {Greffet}},\ }\href {\doibase
  10.1016/j.surfrep.2004.12.002} {\bibfield  {journal} {\bibinfo  {journal}
  {Surface Science Reports}\ }\textbf {\bibinfo {volume} {57}},\ \bibinfo
  {pages} {59} (\bibinfo {year} {2005})}\BibitemShut {NoStop}%
\bibitem [{\citenamefont {Basu}\ \emph {et~al.}(2009)\citenamefont {Basu},
  \citenamefont {Zhang},\ and\ \citenamefont {Fu}}]{Basu2009}%
  \BibitemOpen
  \bibfield  {author} {\bibinfo {author} {\bibfnamefont {S.}~\bibnamefont
  {Basu}}, \bibinfo {author} {\bibfnamefont {Z.~M.}\ \bibnamefont {Zhang}}, \
  and\ \bibinfo {author} {\bibfnamefont {C.~J.}\ \bibnamefont {Fu}},\ }\href
  {\doibase 10.1002/er.1607} {\bibfield  {journal} {\bibinfo  {journal}
  {International Journal of Energy Research}\ }\textbf {\bibinfo {volume}
  {33}},\ \bibinfo {pages} {1203} (\bibinfo {year} {2009})}\BibitemShut
  {NoStop}%
\bibitem [{\citenamefont {Jacobs}\ \emph {et~al.}(2017)\citenamefont {Jacobs},
  \citenamefont {Morgan},\ and\ \citenamefont {Booske}}]{Jacobs2017}%
  \BibitemOpen
  \bibfield  {author} {\bibinfo {author} {\bibfnamefont {R.}~\bibnamefont
  {Jacobs}}, \bibinfo {author} {\bibfnamefont {D.}~\bibnamefont {Morgan}}, \
  and\ \bibinfo {author} {\bibfnamefont {J.}~\bibnamefont {Booske}},\ }\href
  {\doibase 10.1063/1.5006029} {\bibfield  {journal} {\bibinfo  {journal} {APL
  Materials}\ }\textbf {\bibinfo {volume} {5}},\ \bibinfo {pages} {116105}
  (\bibinfo {year} {2017})}\BibitemShut {NoStop}%
\bibitem [{\citenamefont {Van Der~Maas}\ \emph {et~al.}(1985)\citenamefont {Van
  Der~Maas}, \citenamefont {Huguenin},\ and\ \citenamefont
  {Gasparov}}]{VanDerMaas1985}%
  \BibitemOpen
  \bibfield  {author} {\bibinfo {author} {\bibfnamefont {J.}~\bibnamefont {Van
  Der~Maas}}, \bibinfo {author} {\bibfnamefont {R.}~\bibnamefont {Huguenin}}, \
  and\ \bibinfo {author} {\bibfnamefont {V.~A.}\ \bibnamefont {Gasparov}},\
  }\href {\doibase 10.1088/0305-4608/15/11/006} {\bibfield  {journal} {\bibinfo
   {journal} {Journal of Physics F}\ }\textbf {\bibinfo {volume} {15}},\
  \bibinfo {pages} {L271} (\bibinfo {year} {1985})}\BibitemShut {NoStop}%
\bibitem [{\citenamefont {Roberts}(1959)}]{Roberts1959}%
  \BibitemOpen
  \bibfield  {author} {\bibinfo {author} {\bibfnamefont {S.}~\bibnamefont
  {Roberts}},\ }\href {\doibase 10.1103/PhysRev.114.104} {\bibfield  {journal}
  {\bibinfo  {journal} {Physical Review}\ }\textbf {\bibinfo {volume} {114}},\
  \bibinfo {pages} {104} (\bibinfo {year} {1959})}\BibitemShut {NoStop}%
\bibitem [{\citenamefont {Schwede}\ \emph {et~al.}(2010)\citenamefont
  {Schwede}, \citenamefont {Bargatin}, \citenamefont {Riley}, \citenamefont
  {Hardin}, \citenamefont {Rosenthal}, \citenamefont {Sun}, \citenamefont
  {Schmitt}, \citenamefont {Pianetta}, \citenamefont {Howe}, \citenamefont
  {Shen},\ and\ \citenamefont {Melosh}}]{Schwede2010}%
  \BibitemOpen
  \bibfield  {author} {\bibinfo {author} {\bibfnamefont {J.~W.}\ \bibnamefont
  {Schwede}}, \bibinfo {author} {\bibfnamefont {I.}~\bibnamefont {Bargatin}},
  \bibinfo {author} {\bibfnamefont {D.~C.}\ \bibnamefont {Riley}}, \bibinfo
  {author} {\bibfnamefont {B.~E.}\ \bibnamefont {Hardin}}, \bibinfo {author}
  {\bibfnamefont {S.~J.}\ \bibnamefont {Rosenthal}}, \bibinfo {author}
  {\bibfnamefont {Y.}~\bibnamefont {Sun}}, \bibinfo {author} {\bibfnamefont
  {F.}~\bibnamefont {Schmitt}}, \bibinfo {author} {\bibfnamefont
  {P.}~\bibnamefont {Pianetta}}, \bibinfo {author} {\bibfnamefont {R.~T.}\
  \bibnamefont {Howe}}, \bibinfo {author} {\bibfnamefont {Z.-X.}\ \bibnamefont
  {Shen}}, \ and\ \bibinfo {author} {\bibfnamefont {N.~A.}\ \bibnamefont
  {Melosh}},\ }\href {\doibase 10.1038/nmat2814} {\bibfield  {journal}
  {\bibinfo  {journal} {Nature Materials}\ }\textbf {\bibinfo {volume} {9}},\
  \bibinfo {pages} {762} (\bibinfo {year} {2010})}\BibitemShut {NoStop}%
\bibitem [{\citenamefont {Miskovsky}\ \emph {et~al.}(1993)\citenamefont
  {Miskovsky}, \citenamefont {Park}, \citenamefont {He},\ and\ \citenamefont
  {Cutler}}]{Miskovsky1993}%
  \BibitemOpen
  \bibfield  {author} {\bibinfo {author} {\bibfnamefont {N.~M.}\ \bibnamefont
  {Miskovsky}}, \bibinfo {author} {\bibfnamefont {S.~H.}\ \bibnamefont {Park}},
  \bibinfo {author} {\bibfnamefont {J.}~\bibnamefont {He}}, \ and\ \bibinfo
  {author} {\bibfnamefont {P.~H.}\ \bibnamefont {Cutler}},\ }\href {\doibase
  10.1116/1.586685} {\bibfield  {journal} {\bibinfo  {journal} {Journal of
  Vacuum Science {\&} Technology B}\ }\textbf {\bibinfo {volume} {11}},\
  \bibinfo {pages} {366} (\bibinfo {year} {1993})}\BibitemShut {NoStop}%
\bibitem [{\citenamefont {Fisher}\ and\ \citenamefont
  {Walker}(2002)}]{Fisher2002c}%
  \BibitemOpen
  \bibfield  {author} {\bibinfo {author} {\bibfnamefont {T.~S.}\ \bibnamefont
  {Fisher}}\ and\ \bibinfo {author} {\bibfnamefont {D.~G.}\ \bibnamefont
  {Walker}},\ }\href {\doibase 10.1115/1.1494091} {\bibfield  {journal}
  {\bibinfo  {journal} {Journal of Heat Transfer}\ }\textbf {\bibinfo {volume}
  {124}},\ \bibinfo {pages} {954} (\bibinfo {year} {2002})}\BibitemShut
  {NoStop}%
\bibitem [{\citenamefont {Smith}\ \emph {et~al.}(2006)\citenamefont {Smith},
  \citenamefont {Nemanich},\ and\ \citenamefont {Bilbro}}]{Smith2006}%
  \BibitemOpen
  \bibfield  {author} {\bibinfo {author} {\bibfnamefont {J.~R.}\ \bibnamefont
  {Smith}}, \bibinfo {author} {\bibfnamefont {R.~J.}\ \bibnamefont {Nemanich}},
  \ and\ \bibinfo {author} {\bibfnamefont {G.~L.}\ \bibnamefont {Bilbro}},\
  }\href {\doibase 10.1016/j.diamond.2005.12.057} {\bibfield  {journal}
  {\bibinfo  {journal} {Diamond and Related Materials}\ }\textbf {\bibinfo
  {volume} {15}},\ \bibinfo {pages} {870} (\bibinfo {year} {2006})}\BibitemShut
  {NoStop}%
\bibitem [{\citenamefont {Jensen}\ \emph {et~al.}(2006)\citenamefont {Jensen},
  \citenamefont {Lau},\ and\ \citenamefont {Jordan}}]{Jensen2006}%
  \BibitemOpen
  \bibfield  {author} {\bibinfo {author} {\bibfnamefont {K.~L.}\ \bibnamefont
  {Jensen}}, \bibinfo {author} {\bibfnamefont {Y.~Y.}\ \bibnamefont {Lau}}, \
  and\ \bibinfo {author} {\bibfnamefont {N.}~\bibnamefont {Jordan}},\ }\href
  {\doibase 10.1063/1.2197605} {\bibfield  {journal} {\bibinfo  {journal}
  {Applied Physics Letters}\ }\textbf {\bibinfo {volume} {88}},\ \bibinfo
  {pages} {1} (\bibinfo {year} {2006})}\BibitemShut {NoStop}%
\bibitem [{\citenamefont {Bernardi}\ \emph {et~al.}(2016)\citenamefont
  {Bernardi}, \citenamefont {Milovich},\ and\ \citenamefont
  {Francoeur}}]{Bernardi2016a}%
  \BibitemOpen
  \bibfield  {author} {\bibinfo {author} {\bibfnamefont {M.~P.}\ \bibnamefont
  {Bernardi}}, \bibinfo {author} {\bibfnamefont {D.}~\bibnamefont {Milovich}},
  \ and\ \bibinfo {author} {\bibfnamefont {M.}~\bibnamefont {Francoeur}},\
  }\href {\doibase 10.1038/ncomms12900} {\bibfield  {journal} {\bibinfo
  {journal} {Nature Communications}\ }\textbf {\bibinfo {volume} {7}},\
  \bibinfo {pages} {1} (\bibinfo {year} {2016})}\BibitemShut {NoStop}%
\bibitem [{\citenamefont {DeSutter}\ \emph {et~al.}(2019)\citenamefont
  {DeSutter}, \citenamefont {Tang},\ and\ \citenamefont
  {Francoeur}}]{DeSutter2019}%
  \BibitemOpen
  \bibfield  {author} {\bibinfo {author} {\bibfnamefont {J.}~\bibnamefont
  {DeSutter}}, \bibinfo {author} {\bibfnamefont {L.}~\bibnamefont {Tang}}, \
  and\ \bibinfo {author} {\bibfnamefont {M.}~\bibnamefont {Francoeur}},\ }\href
  {\doibase 10.1038/s41565-019-0483-1} {\bibfield  {journal} {\bibinfo
  {journal} {Nature Nanotechnology}\ } (\bibinfo {year} {2019}),\
  10.1038/s41565-019-0483-1}\BibitemShut {NoStop}%
\bibitem [{\citenamefont {Ghashami}\ \emph {et~al.}(2018)\citenamefont
  {Ghashami}, \citenamefont {Geng}, \citenamefont {Kim}, \citenamefont
  {Iacopino}, \citenamefont {Cho},\ and\ \citenamefont {Park}}]{Ghashami2018b}%
  \BibitemOpen
  \bibfield  {author} {\bibinfo {author} {\bibfnamefont {M.}~\bibnamefont
  {Ghashami}}, \bibinfo {author} {\bibfnamefont {H.}~\bibnamefont {Geng}},
  \bibinfo {author} {\bibfnamefont {T.}~\bibnamefont {Kim}}, \bibinfo {author}
  {\bibfnamefont {N.}~\bibnamefont {Iacopino}}, \bibinfo {author}
  {\bibfnamefont {S.~K.}\ \bibnamefont {Cho}}, \ and\ \bibinfo {author}
  {\bibfnamefont {K.}~\bibnamefont {Park}},\ }\href {\doibase
  10.1103/PhysRevLett.120.175901} {\bibfield  {journal} {\bibinfo  {journal}
  {Physical Review Letters}\ }\textbf {\bibinfo {volume} {120}},\ \bibinfo
  {pages} {175901} (\bibinfo {year} {2018})}\BibitemShut {NoStop}%
\end{thebibliography}%

\clearpage

\section*{\large Figure Captions}
\noindent \textbf{Fig. 1}. (a) A schematic diagram of the thermionic potential profile across a submicron vacuum gap. The electric field between the electrodes that shapes the potential profile is caused by the work function difference, the voltage drop over the load and the image-charge potential. The resulting electric current across the vacuum gap occurs classically (thermionic) and quantum mechanically (electron tunneling). (b) TEC device schematic. In addition to charge transport due to electrons, heat is transferred across a vacuum space by electrons and thermal radiation, partly converted to electric power and remaining rejected to a room-temperature reservoir. \vspace{12 pt}

\noindent \textbf{Fig. 2}. Potential barrier profiles, $W(x)$, and the maximum potential $W_{\textrm{max}}$ for three vacuum gaps ($d=500$ nm, 3 $\mu$m, and 10 $\mu$m) when $T_E=1575$ K ($\Phi_E=2.10$ eV) and $T_C=1000$ K ($\Phi_C=1.33$ eV). (a) Ideal potential profiles at different load voltages, where the flat band voltage ($V_\mathrm{FB}$) is 0.8 V.  (b),(c) Realistic potential profiles for $d = 10$ $\mu$m and $d = 3$ $\mu$m, respectively, present the increase of the potential barrier due to the space charge effect. In the figures, $V_B$ and $V_S$ denote the Boltzmann voltage and saturation voltage, respectively. (d) Potential profiles for $d = 500$ nm, which is slightly lower than the ideal profiles due to image charge effect. (e) The relation between $W_\textrm{max}$ and the load voltage for different gap distances. The locations of $W_\textrm{max}$ are marked with solid square points in (b-d).  \vspace{12 pt}

\noindent \textbf{Fig. 3}. The net thermionic ($J_\mathrm{TE}$; top) and electron tunneling ($J_\mathrm{QE}$; bottom) current densities across three vacuum gaps ($d=$ 500 nm, 3 $\mu$m and 10 $\mu$m) for the same operation condition as Fig. 2. $J_\mathrm{TE}$ curves for $d = 3$ $\mu$m and 10 $\mu$m are smaller than the ideal case due to space charge accumulation. 
For $d = 500$ nm, however, electrons are accelerated due to interactions with image charges to provide a current density larger than the ideal case. When compared to $J_\mathrm{TE}$, $J_\mathrm{QE}$ is about two orders of magnitude smaller and may not contribute to overall power generation. The solid square marks on the $J_\mathrm{TE}$ curves denote the maximum power operation points. 

\noindent \textbf{Fig. 4}. The gap-dependent TEC performance characteristics operating at constant emitter and collector temperatures ($T_E=1575$ K and $T_C = 1000$ K). (a) The maximum power density (black dashed) due to thermionic emission (red) and electron tunneling (blue). (b) The total heat flux across the vacuum gap (black) due to thermionic emission (red), electron tunneling (blue) and thermal radiation (green). (c) The resulting energy conversion efficiency and (d) the required convective heat transfer coefficient of a cooling fluid to maintain a constant $T_\textrm{C} = 1000$ K demonstrate the optimum gap distance for the given operation condition is in the range of 300 nm $\lesssim d \lesssim$ 1 $\mu$m. \vspace{12 pt}

\noindent \textbf{Fig. 5}. The performance characteristics of the TEC device operating at the same operation condition as Fig. 4 except constant convection heat transfer coefficient $h_\infty = 1000$ W/m$^2$-K. (a) The collector temperature and corresponding work function, (b) total heat flux, (c) total power output, and (d) energy conversion efficiency as a function of the vacuum gap provide the consistent optimum vacuum gap range. The red hatched areas represent unrealistic TEC characteristics due to $\Phi_\textrm{C}<1$ eV. \vspace{12 pt}

\noindent \textbf{Fig. 6}. The increase in (a) the maximum power output and (b) efficiency by combining a TEC cycle with a bottom-cycle heat engine with 30\% efficiency. Heat rejection from the collector, ordinarily wasted, is now harnessed to generate additional power in the bottom cycle to improve the system efficiency. The efficiency reaches $\sim$48\% at $d\sim 500$ nm, and for $d<100$ nm a power density is generated greater than 100 W/cm$^2$ at an efficiency larger than 30\%.  \vspace{12 pt}

\noindent \textbf{Fig. 7}. (a) The maximum power output, (b) energy conversion efficiency and (c) required heat transfer coefficient of a cooling fluid as a function of the linear field enhancement factor $\beta$. As $\beta$ increases by localizing the electric field at protrusion apexes of engineered electrode surfaces, the power output is greatly enhanced due to stronger field-induced charge acceleration. However, more thermionic heat is transferred to the collector, resulting in the decrease of the energy conversion efficiency and the increase of the required heat transfer coefficient to maintain the collector temperature. 

\clearpage

\clearpage

\clearpage

\clearpage


\clearpage

\clearpage

\clearpage

\clearpage

\clearpage

\end{document}